%% file: main.tex
\documentclass[10pt,journal,compsoc]{IEEEtran}
\input{_header}
\usepackage{hyperref}

\begin{document}

\title{Iterative Window Mean Filter: Thwarting Diffusion-based Adversarial Purification}

\input{_authors}
%\input{_author_reimbursment}

\markboth{IEEE Transactions on Dependable and Secure Computing,~Vol.~X, No.~X, September~2024}
{Hanrui Wang, Ruoxi Sun,
\MakeLowercase{\textit{(et al.)}:
Iterative Window Mean Filter: Thwarting Diffusion-based Adversarial Purification}
}

\maketitle

\begin{abstract}
Face authentication systems have brought significant convenience and advanced developments, yet they have become unreliable due to their sensitivity to inconspicuous perturbations, such as adversarial attacks. Existing defenses often exhibit weaknesses when facing various attack algorithms and adaptive attacks or compromise accuracy for enhanced security. To address these challenges, we have developed a novel and highly efficient non-deep-learning-based image filter called the Iterative Window Mean Filter (IWMF) and proposed a new framework for adversarial purification, named IWMF-Diff, which integrates IWMF and denoising diffusion models. These methods can function as pre-processing modules to eliminate adversarial perturbations without necessitating further modifications or retraining of the target system. We demonstrate that our proposed methodologies fulfill four critical requirements: preserved accuracy, improved security, generalizability to various threats in different settings, and better resistance to adaptive attacks. This performance surpasses that of the state-of-the-art adversarial purification method, DiffPure. Our code is released at \href{https://github.com/azrealwang/iwmfdiff}{https://github.com/azrealwang/iwmfdiff}.
% {\let\thefootnote\relax\footnote{{$^{\ast}$~Corresponding authors.}}}
\end{abstract}


\section{Introduction}
\label{motivation_df}
\begin{figure}[t]
\setlength{\abovecaptionskip}{0.1cm}
\setlength{\belowcaptionskip}{0.1cm}
\centering
\includegraphics[width=\linewidth]{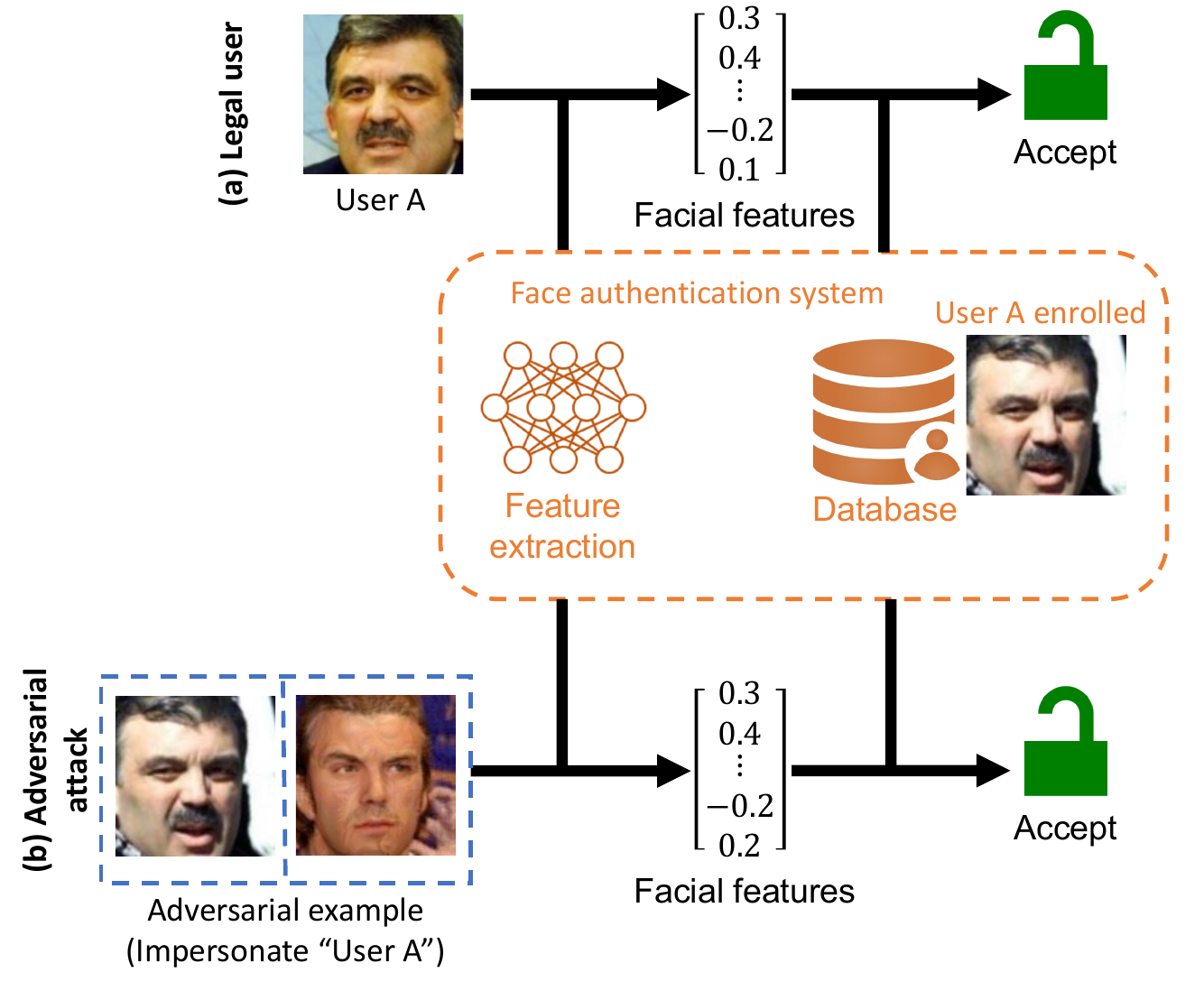}
\caption{Adversarial attack against face authentication. (b) represents an impersonation attack.}
\label{fig_adv_attack}
\end{figure}

\IEEEPARstart{D}{eep} learning has made significant strides in security applications, such as face authentication, achieving impressive performance. However, adversarial attacks have emerged as a major threat to the authentication security. In the context of the face authentication, an adversarial attack refers to a technique that leverages a deceptive input (\ie adversarial example~\cite{Kur16}) to mislead the decision from rejection to acceptance, as illustrated in Figure~\ref{fig_adv_attack}. Such attacks can result in unauthorized access to authentication systems. Consequently, defenses against adversarial attacks are essential to secure security systems.

While many adversarial defenses have been proposed, they often suffer one or more problems, rendering them impractical for real-world implementation against sophisticated attackers. For instance, detection models~\cite{grosse2017statistical} are typically trained on adversarial examples from specific attacks, making it challenging to detect other attacks. Furthermore, such defenses are often limited to binary classification, distinguishing between adversarial and non-adversarial inputs. Robustness optimization techniques, such as adversarial training~\cite{goodfellow2014explaining}, require vast amounts of data to train the model and still struggle to mitigate unexposed and adaptive attacks. Although the traditional adversarial purification methods, such as randomized blurring~\cite{wang2017adversary,XuW171,li2017adversarial,graese2016assessing,guo2018countering}, have shown potential in enhancing generalization resistance, they usually accomplish this by sacrificing the accuracy of the system.

Adversarial purification by generative models has emerged as one of the most viable defense strategies and is currently attracting a lot of research attention. This is because it has been observed to be effective in countering both various attack algorithms and adaptive attacks. Specifically, adaptive attacks are designed to undermine the defense strategy and often have full knowledge of both the deep learning model and its associated defense strategies~\cite{carlini2017adversarial, wang2021similarity}. The ability to resist various attacks is considered one of the most difficult challenges of adversarial defenses~\cite{ren2022perturbation, nie2022DiffPure}. To achieve adversarial defense, auto-encoder-based approaches~\cite{Buc18,zhou2020manifold,vahdat2020nvae,ren2022perturbation} adopt Variational Autoencoder (VAE)~\cite{kingma2013auto} to train more robust feature extractors. Meanwhile, diffusion-based defenses~\cite{nie2022DiffPure,wang2022guided,blau2022threat,sun2022pointdp,wu2022guided} reverse the diffusion process of diffusion models to generate clean images, and perturbations are concealed by Gaussian noise. Among all these types of defenses, DiffPure~\cite{nie2022DiffPure} has achieved state-of-the-art performance.

However, there exists critical defects in auto-encoder- and diffusion-based defenses. Concerning auto-encoder-based defenses, their performance is less satisfactory compared with diffusion-based methods, and they are ineffective against adaptive attacks. For diffusion-based defenses, there are several issues to consider. Firstly, due to the complexity of the diffusion models, these defenses may suffer from computational exhaustion~\cite{nie2022DiffPure}. Secondly, all existing diffusion-based defenses rely on Gaussian noise to conceal adversarial perturbations, which is shown to be impractical for classifying adversarial examples as their true labels because the noise changes or obscures facial features. Thirdly, diffusion-based defenses in a fixed setting may be unable to defend against attacks with large perturbation sizes, because the fixed Gaussian noise may not be able to conceal severe adversarial perturbations. In other words, a necessary condition for success by the state-of-the-art defense is having the knowledge of the attack settings (\eg perturbation size), making it less applicable to other settings. Finally, diffusion-based defenses are not effective against specific black-box adversarial attacks and adaptive attacks. Additionally, there is currently no widely-accepted criteria to evaluate adversarial defenses or to make comprehensive comparisons amongst them. Consequently, we propose four requirements for evaluating an ideal adversarial defense, as follows:

\begin{itemize}[leftmargin=*]
\item \textit{Accuracy.} The accuracy on genuine images, \ie non-adversarial images, must be preserved.
\item \textit{Security.} An effective defense should be robust against adversarial examples in two ways: \one~the system should classify adversarial examples as their true labels, which resists indiscriminate and data poisoning attacks, such as backdoor attacks; or \two~the defense system should NOT classify adversarial examples as the target labels, which resists targeted attacks, such as impersonation attacks.
\item \textit{Generalization.} An adversarial defense should be able to be generally applied to various threat models, \eg white/gray/black-box attacks, and effective against various attack algorithms in different settings (\eg~in a larger perturbation size).
\item \textit{Resistance against adaptive attacks.} An ideal defense shows resistance against adaptive attacks. This is the most challenging task, yet overlooked by many papers.
\end{itemize}

On top of these, we design adversarial defenses to meet the four specified requirements and evaluate the proposed methodologies against these more comprehensive criteria. First, we propose an innovative image filter, the Iterative Window Mean Filter (IWMF), which is derived from the classic mean filter. IWMF is a non-deep learning-based method used to conceal adversarial perturbations, which resolves the efficiency issue facing deep learning-based defenses (such as auto-encoder and diffusion). IWMF strengthens security while simultaneously preserves the accuracy on genuine inputs during verification. Taking advantage of IWMF, we propose an image pre-processing framework called IWMF-Diff, illustrated in Figure~\ref{fig_pipeline}. In IWMF-Diff, both genuine images and adversarial examples are blurred using IWMF and then restored using Denoising Diffusion Restoration Models (DDRM)~\cite{kawar2022denoising}. This approach mitigates the decline in the genuine image authentication accuracy that comes along with using IWMF on its own, and is more robust against various attack algorithms \cite{goodfellow2014explaining,madry2018towards,carlini2017towards,wang2021similarity,croce2020reliable,lee2023robust,kurakin2018adversarial,dong2019evading,xie2019improving,yang2020robfr,andriushchenko2020square} in different settings, including larger perturbation sizes. The proposed methods also show better resistance against adaptive attacks compared with pure auto-encoder or diffusion-based methods, by decreasing the attack success rate from 99.4\% for DiffPure to 77.4\% for our IWMF-DIFF. Regarding performance, IWMF outperforms all other blurring strategies, including Gaussian noise, and even individually delivers comparable performance with the state-of-the-art diffusion-based defense~\cite{nie2022DiffPure}. IWMF-Diff satisfies all four requirements for ideal adversarial defenses and outperforms the state-of-the-art defense. In summary, we have made the following contributions:

\begin{itemize}[leftmargin=*] 
\item We defined four requirements for ideal adversarial defenses that can be used as standards to evaluate new defenses. We discussed the necessity of these requirements and presented experimental evidence of their importance from a security standpoint. 
\item We conducted a detailed investigation and found that adversarial defenses that use auto-encoder and Gaussian-based diffusion models, despite still receiving considerable research attention, are impractical in real-world scenarios because they cannot meet all four requirements. 
\item We proposed an innovative non-deep-learning-based image filter, IWMF, which can erase adversarial perturbations. IWMF individually outperforms other blurring strategies and the latest auto-encoder-based adversarial purification methods. It does not require training or high-performance computing and can deliver comparable performance with the state-of-the-art diffusion-based adversarial defense. Furthermore, IWMF is exceptionally efficient, making it suitable for real-time tasks. 
\item We proposed a diffusion-based image processing framework, IWMF-Diff, for adversarial purification, which exploits IWMF. IWMF-Diff can be used as a pre-processing module for any system without further modification or training. IWMF-Diff satisfies all four requirements and outperforms the state-of-the-art adversarial defense. 
\end{itemize}

\section{Related Work}
%In this section, we introduce the background and related works of our study.

\subsection{Adversarial Defenses}
Existing adversarial defenses can be divided into four categories: \one\ gradient masking, \two\ adversarial example detection, \three\ robustness optimization, and \four\ adversarial purification.

Most adversarial attacks rely on the gradient information of the classifier. Gradient masking aims to hide this gradient information from adversaries. It has been shown to be effective in confusing attacks but has been replaced by adversarial purification. A distillation method was first proposed in~\cite{Hin15} and later reformulated by~\cite{Pap16}. Distillation aims to reduce the size of deep learning models and can be used against adversarial attacks~\cite{szegedy2014intriguing,goodfellow2014explaining,moosavi2016deepfool}. The gradient information can also be hidden by randomizing the models~\cite{dhillonstochastic,xiemitigating}, making it difficult for attackers to determine which model is being used. However, gradient masking can only ``confuse'' attacks, rather than eliminate them entirely, and can be counterattacked by methods such as~\cite{carlini2017towards,athalye2018obfuscated}. The ``mask'' can also be overwhelmed by a surrogate classifier whose gradient is known to the attacker~\cite{papernot2017practical,carlini2019evaluating}.

Robustness optimization involves attempts to classify adversarial examples as their true labels by training a more robust model. Szegedy \etal~\cite{szegedy2014intriguing} were the first to address this idea by regularizing the training process to increase the stability of the output. Building on~\cite{szegedy2014intriguing}, Cisse \etal~\cite{cisse2017parseval} and Miyato \etal~\cite{miyato2015distributional} constrained the instability of the Lipschitz constant (a bound on the rate of change of the objective function~\cite{jones1993lipschitzian}). Gu and Rigazio~\cite{gu2014towards} proposed a deep contractive network to regularize the partial derivatives at each layer.
Goodfellow \etal~\cite{goodfellow2014explaining} proposed adversarial training by introducing FGSM adversarial examples, along with their ground-truth labels, into the training dataset. Building on this work, Madry \etal~\cite{madry2018towards} expanded the scope of adversarial training techniques by using more types of adversarial examples to improve model robustness and enhance generalization against a variety of attacks.
Shafahi \etal~\cite{shafahi2019adversarial} proposed a technique to enhance the efficiency of training large-scale datasets by reusing backward pass computations, which was further improved in \cite{zhang2019you}. However, while these methods improve model robustness, they are unlikely to be effective against various attack algorithms. In other words, adversarial examples that are not included in the training process, such as those generated by Xu \etal~\cite{xu2020adversarial}, Tramer \etal~\cite{tramerensemble}, and Miller \etal~\cite{miller2020adversarial}, may not be mitigated through the use of robustness optimization.

Since increasing model robustness is challenging, some studies focus on detecting adversarial examples from benign images. Grosse \etal~\cite{grosse2017statistical} trained an auxiliary model for detection, assigning an extra label to represent all adversarial examples. They also observed that the distributions between benign images and adversarial examples differ, and used the maximum mean discrepancy (MMD) test~\cite{gretton2012kernel} to examine this discrepancy. Hendrycks and Gimpel~\cite{hendrycks2016early} distinguished between adversarial examples and benign images using principal component analysis, while Feinman \etal~\cite{feinman2017detecting} detected adversarial examples by identifying inconsistencies in classification results from randomly-selected deep learning models. However, adversarial example detection is also limited in its effectiveness against various attacks and can be countered~\cite{carlini2017adversarial}.

Adversarial purification aims to cover the perturbations from adversarial examples. One early approach involved distorting the images. For example, Wang \etal~\cite{wang2017adversary} nullified pixels irregularly, while Xu \etal~\cite{XuW171} used quantization to cover small perturbations. An alternative method employed convolutional filter statistics to distort images, as used by Li and Li~\cite{li2017adversarial}. Graese \etal~\cite{graese2016assessing} proposed an image cropping technique to destroy adversarial examples, while Guo \etal~\cite{guo2018countering} pre-processed input, including adversarial examples, using a non-differentiable transformation. However, while these techniques can help to combat adversarial examples, they often lead to a significant trade-off in classification accuracy for genuine images due to the distortion.

A more promising approach to adversarial purification involves generating replacements for images and feeding these regenerated images to deep learning models. PixelDefend, proposed by Song \etal~\cite{song2018pixeldefend}, is based on generative adversarial network (GAN). Another GAN-based method for adversarial defense is Defense-GAN, devised by Samangouei \etal~\cite{samangouei2019defense}. In an alternative approach, Buckman \etal~\cite{Buc18} used thermometer encoding to regenerate images, inspiring the work of Zhou \etal~\cite{zhou2020manifold} and Ren \etal~\cite{ren2022perturbation}, who developed individual image generating models using auto-encoder technology~\cite{kingma2013auto}. GAN- and auto-encoder-based adversarial purification techniques can be vulnerable to adaptive attacks, leading to unsatisfactory defense performance. As such, there is growing interest in using diffusion models~\cite{ho2020denoising} for adversarial purification~\cite{nie2022DiffPure, wang2022guided, blau2022threat, sun2022pointdp, wu2022guided}. Recently, Nie \etal~\cite{nie2022DiffPure} introduced a state-of-the-art diffusion-based purification technique called DiffPure. However, while diffusion-based adversarial purification techniques, such as DiffPure~\cite{nie2022DiffPure}, have achieved impressive results, they can be computationally expensive~\cite{nie2022DiffPure}. Additionally, all existing diffusion-based defenses employ Gaussian noise to cover perturbations, which limits their effectiveness in classifying adversarial examples correctly, defending black-box adversarial attacks, and defending against general attacks with large perturbations. Furthermore, these methods often perform poorly against adaptive attacks.

\subsection{Denoising Diffusion Models}
Denoising diffusion models are a type of latent variable models that use variational inference to train Markov chains and learn the latent structure of a dataset. The models simulate the way in which data points diffuse through the latent space, and this forward process can be reversed to denoise images that have been blurred with Gaussian noise. Adversarial defenses utilize denoising diffusion models to purify adversarial examples by generating replacements that reduce the perturbations~\cite{nie2022DiffPure, wang2022guided, blau2022threat, sun2022pointdp, wu2022guided}. Denoising Diffusion Probabilistic Models (DDPM)~\cite{ho2020denoising} is one of the most commonly-used denoising diffusion models, but more improved models have since been introduced, such as Denoising Diffusion Implicit Models (DDIM)~\cite{song2020denoising} and DDRM~\cite{kawar2022denoising}. DDPM and DDIM serve as the backbones of DDRM, and the diffusion module used in the proposed IWMF-Diff is derived from DDRM. More details of DDPM, DDIM, and DDRM can be found in Appendix~B.

\begin{figure}[t]
\setlength{\abovecaptionskip}{0.1cm}
\setlength{\belowcaptionskip}{0.1cm}
\centering
\includegraphics[width=\linewidth]{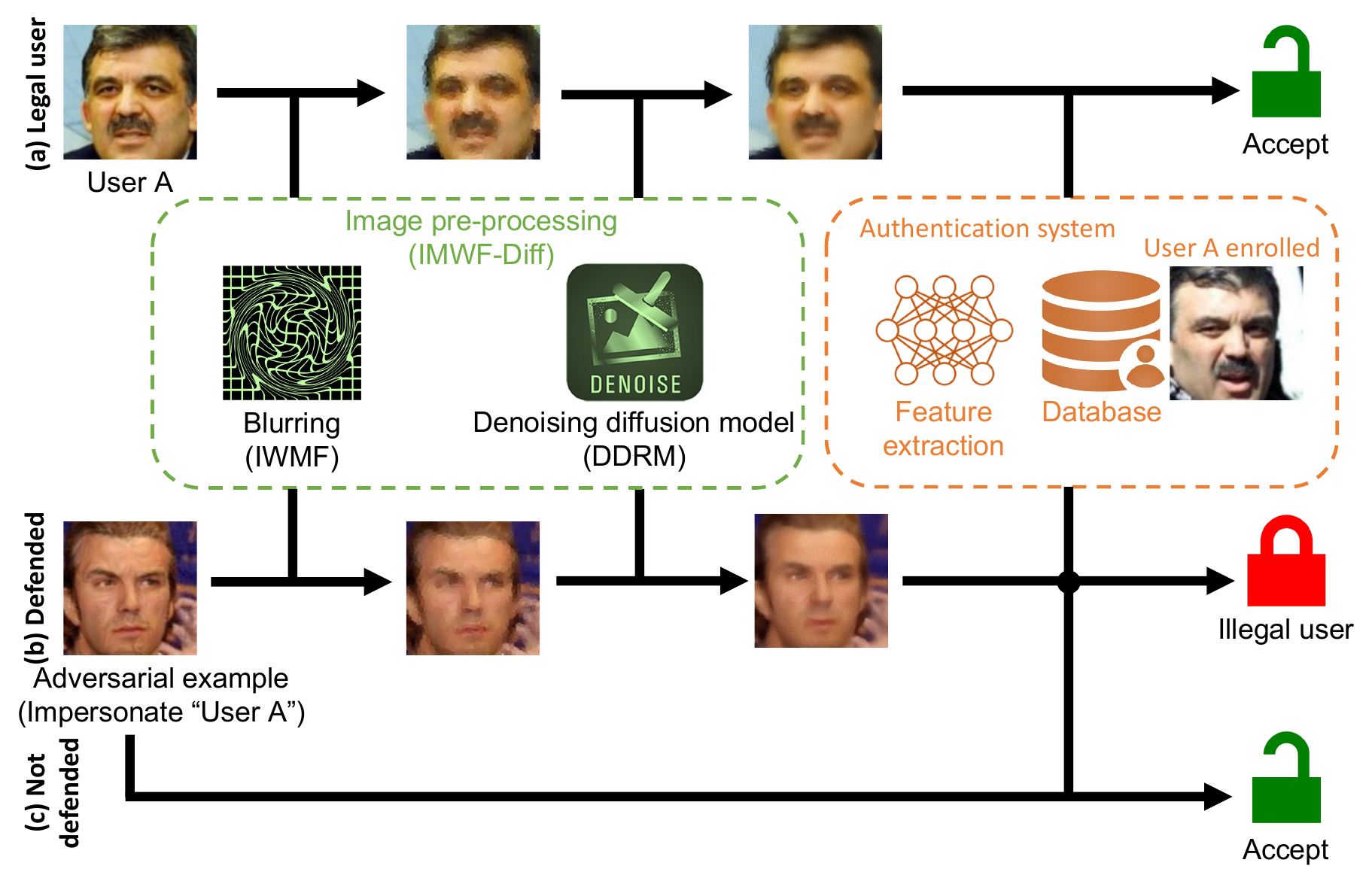}
\caption{Framework of IWMF-Diff.}
\label{fig_pipeline}
\end{figure}

\section{IWMF-Diff Framework}
IWMF-Diff is a pre-processing module designed for adversarial purification before authentication. All inputs, including genuine images and adversarial examples, are first blurred using the proposed image filter IWMF to cover perturbations. The distorted images are then restored by the denoising diffusion model DDRM to enhance robustness for verification. Following these two pre-processing steps, the images are fed into the regular authentication system, \ie the deep learning model, for verification. IWMF-Diff requires no modification to existing modules, such as the feature extractor and database, and can be seamlessly integrated into any system, with the ability to be easily enabled or disabled. Note that while IWMF-Diff can be used for enrollment, this is not discussed in this paper, as the focus is on protecting authentication. The pipeline for IWMF-Diff is illustrated in Figure~\ref{fig_pipeline}.
%%%%%%%%%%%%%%%%%%%%%%%%%%%%%%%%%%
\subsection{Threat Model}
We consider a challenging threat model from the defender's perspective \cite{suciu2018does,thys2019fooling,laidlaw2021perceptual}. Specifically, we assume that the attacker possesses full knowledge of the target system, including the defense strategies in place. In the worst-case scenario, the attacker may execute white-box $L_\infty$-norm adversarial attacks and employ algorithm-specific adaptive attacks to circumvent the defense mechanisms. Conversely, we posit that the defender has no prior knowledge of the attacks, including the attack settings and algorithms. Consequently, the defender must rely on a fixed but generalizable algorithm and settings to mitigate these threats.

\subsection{Iterative Window Mean Filter (IWMF)}
\label{iwmf}
\begin{figure}[t]
\setlength{\abovecaptionskip}{0.1cm}
\setlength{\belowcaptionskip}{0.1cm}
\centering
\includegraphics[width=\linewidth]{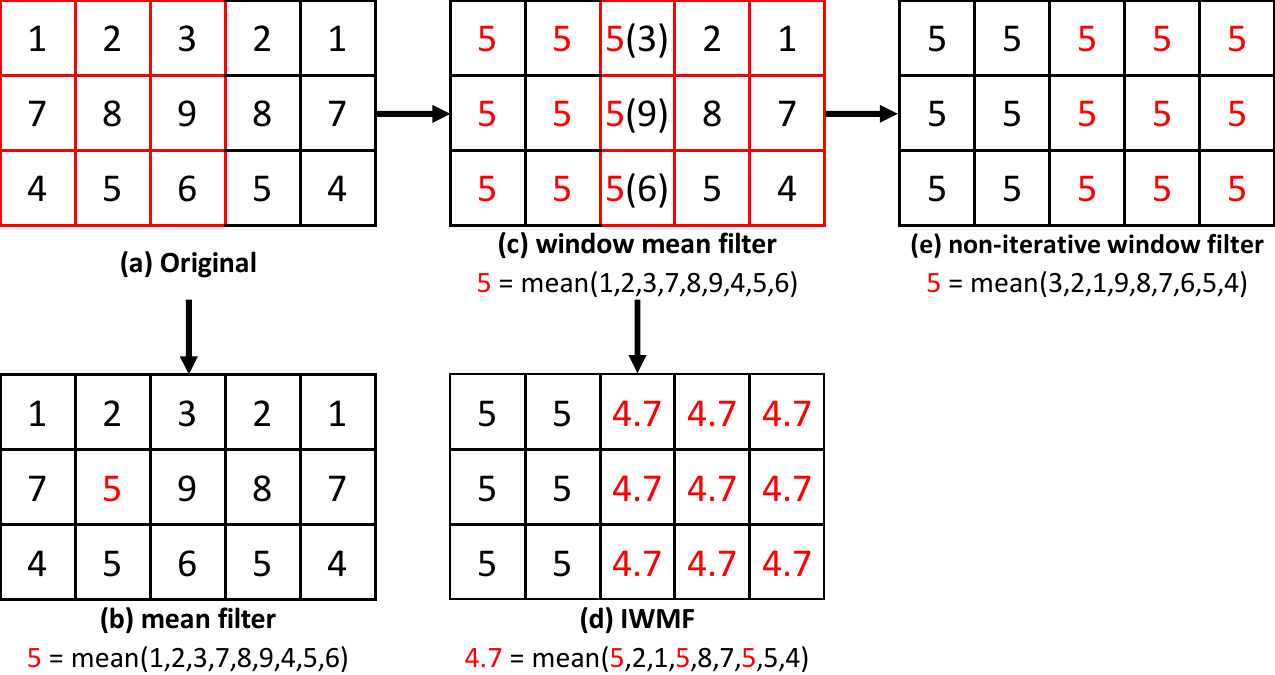}
\caption{Iterative window mean filter (IWMF).}
\label{fig_iwmf}
\end{figure}

Like the classic mean filter, IWMF is designed to reduce noise, specifically adversarial perturbations, by smoothing the image and reducing the amount of intensity variation between pixels. Unlike the traditional mean filter, IWMF enhances the distortion by enlarging the replaced area from a single pixel (Figure~\ref{fig_iwmf}(b)) to the entire window (Figure~\ref{fig_iwmf}(c)). Therefore, IWMF outperforms the traditional mean filter in terms of the robustness against adversarial attacks.

To be more specific, the classic mean filtering process involves computing the average value of the corrupted image $g(x, y)$ in the rectangular window of size $m \times n$, centered at point $(x, y)$. The value at point $(x, y)$ is then replaced by the mean computed using the pixels in the region defined by $S_{xy}$.

\begin{equation}
\label{eq_mean_filter}
\hat{f}(x,y)=\frac{1}{mn}\sum_{(i,j)\in S_{xy}}{g(i,j)}
\end{equation}

In contrast, IWMF replaces all values in the window $S_{xy}$ with the mean value, so:

\begin{equation}
\label{eq_iwmf}
\hat{f}(\forall(i,j)\in S_{xy})=\frac{1}{mn}\sum_{(i,j)\in S_{xy}}{g(i,j)}
\end{equation}

\begin{algorithm}[t]
\caption{Iterative window mean filter (IWMF)}
\label{alg_iwmf}
\begin{algorithmic}[1]
\Require{Face image X, window amount $\lambda$, window size s} 
\Ensure{Processed image Y}
\State \(Y \leftarrow X\) 
\State \(CH,H,W \leftarrow X.shape\)
\State \(Iters = int(\lambda \times H \times W)\) \label{step_lambda}
\For{c in range(CH)}
\For{i in range(Iters)}
\State \((m,n) \leftarrow random\ position\) \label{step_random}\algorithmiccomment{window center}
\State \(window = [m-floor(s/2):m+ceil(s/2),\)\label{step_window}
\Statex \qquad \qquad \qquad \qquad \ \ \(n-floor(s/2):n+ceil(s/2)]\)
\State \(replace = mean(Y[c,window])\)\algorithmiccomment{iterative}\label{step_mean}
\State \(Y(c,window) \leftarrow replace\)\label{step_replace}
\EndFor
\EndFor
\State \Return{Y}
\end{algorithmic}
\end{algorithm}

Moreover, the previous changes of neighbour windows bring new values to next (Figure~\ref{fig_iwmf}(d)). Finally, to ensure the applicability to various types of attacks, each window is randomly selected (Step \ref{step_random} in Algorithm~\ref{alg_iwmf}). The number of windows is determined by a parameter $\lambda$.

\begin{equation}
\label{eq_lambda}
Iters = int(\lambda \times H \times W),
\end{equation}
where $(H,W)$ are the image height and width, respectively.

The complete processing procedure of IWMF is presented in Algorithm~\ref{alg_iwmf}. Window size $s$ denotes the mean calculation and replacement area. $s$ is fixed at 3px in IWMF to gain the best performance. $CH$ represents the image channels. Specifically, the window amount is decided in Step \ref{step_lambda}. Then, for each channel, every window's center position is randomly selected in Step \ref{step_random}, followed by the window chosen in Step \ref{step_window}. Afterwards, the nine original values in the window are replaced by the window's mean value in Steps \ref{step_mean} and \ref{step_replace}. Note that in Step \ref{step_mean}, only if the window is clipped from the continuously optimized image $Y$, it is an iterative filter (\ie IWMF). If the window is from the original input image $X$, it is an non-iterative window mean filter as illustrated in Figure~\ref{fig_iwmf}(e).

The IWMF is specifically designed to fulfill all four requirements of an ideal adversarial defense. 

\begin{figure}[t]
\setlength{\abovecaptionskip}{0.1cm}
\setlength{\belowcaptionskip}{0.1cm}
\centering
\includegraphics[width=\linewidth]{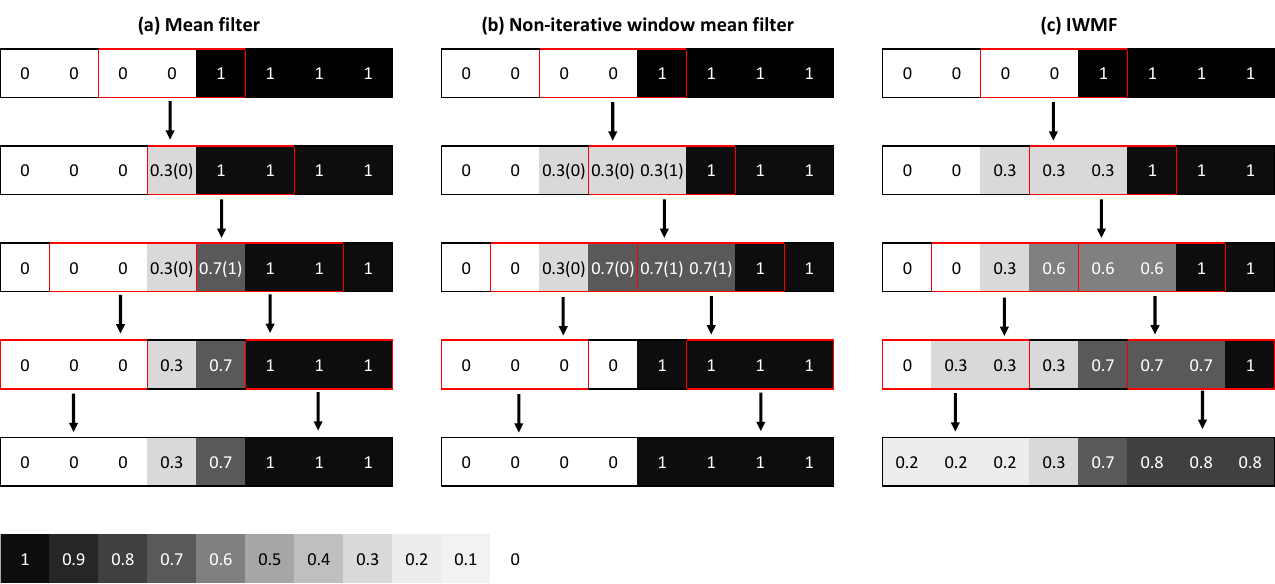}
\caption{The change after processed by IWMF. The edge is smoothed, but still observable for IWMF.}
\label{fig_edge}
\end{figure}

\noindent
\textbf{Accuracy preserving.~}
The mean filter usually smooths the edges between image features, but the edges are still preserved, as illustrated in Figures~\ref{fig_edge}(a) and (b). In contrast, IWMF, as shown in Figure~\ref{fig_edge}(c), reinforces the blurring, resulting in less sharp edges, but still with noticeable color differences at the edges. The preservation of edges helps retain facial features in the filtered image after applying IWMF. The reinforcement of blurring by IWMF does not entirely eliminate feature edges since all pixel value changes are confined within a ``neighbour'' distance, as illustrated in Figure~\ref{fig_iwmf}(d). Furthermore, the number of windows utilized is determined by $\lambda > 0$ and the selection of the windows is achieved randomly. Thus, if the value of $\lambda$ is not sufficiently large, not every pixel in the image will be altered.

\begin{figure}[t]
\setlength{\abovecaptionskip}{0.1cm}
\setlength{\belowcaptionskip}{0.1cm}
\centering
\includegraphics[width=0.8\linewidth]{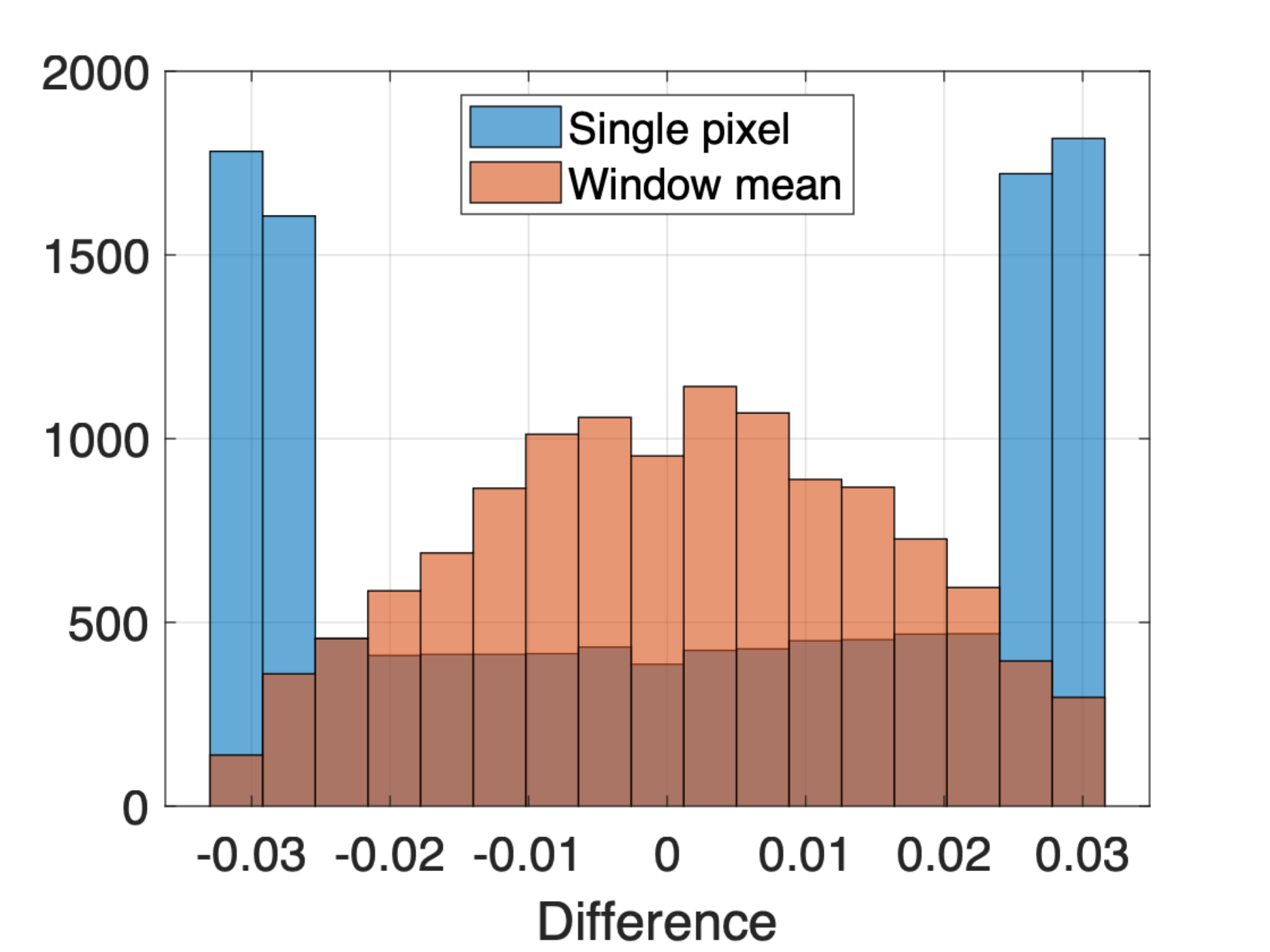}
\caption{The distribution of the differences between the adversarial examples and source images typically exhibit maximum absolute values before defense, but after being processed through IWMF, the differences tend to cluster around zero.}
\label{fig_distribution}
\end{figure}

\begin{figure}[t]
\setlength{\abovecaptionskip}{0.1cm}
\setlength{\belowcaptionskip}{0.1cm}
\centering
\includegraphics[width=0.9\linewidth]{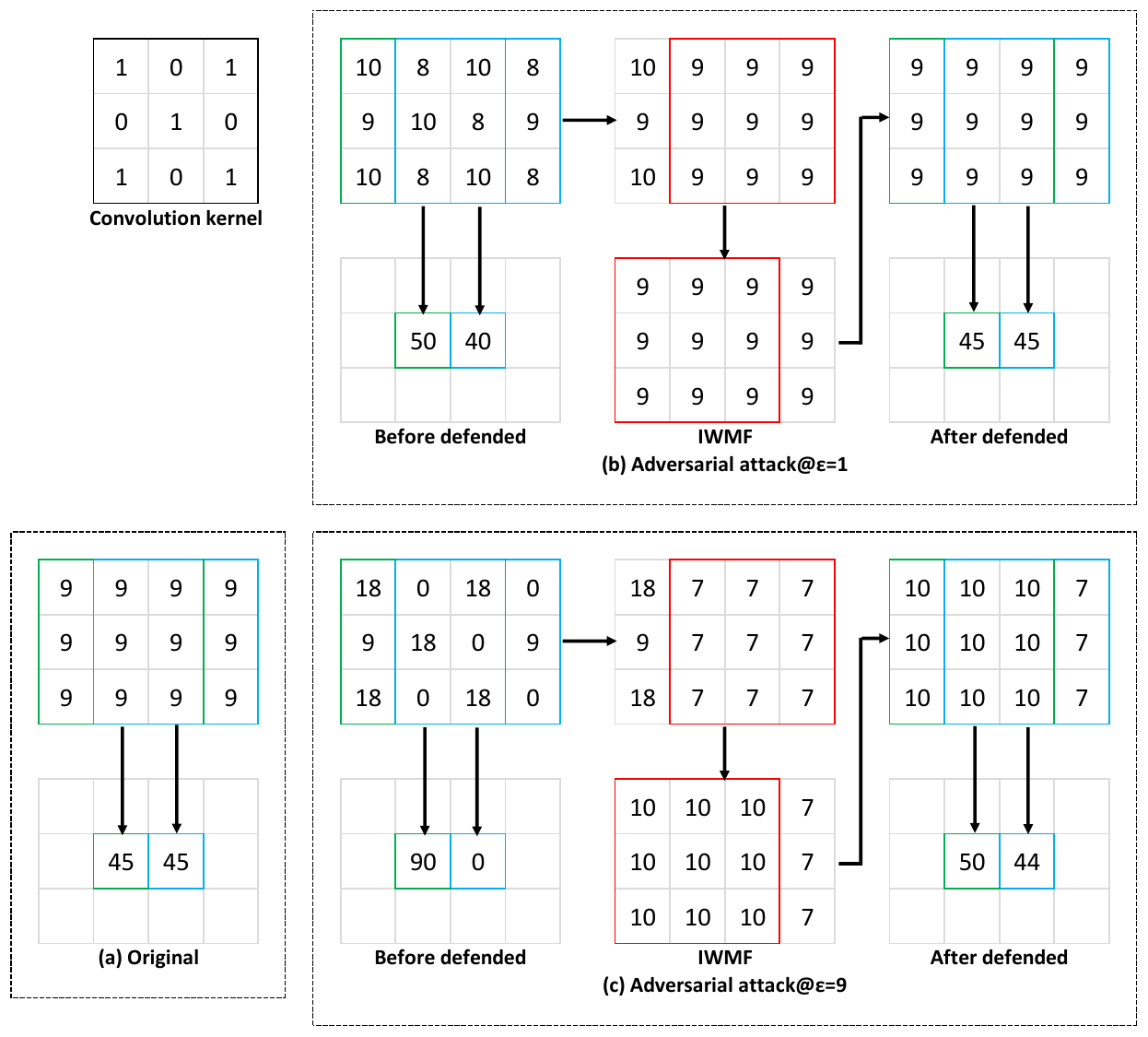}
\caption{The application of IWMF in the image defense process effectively safeguards the first convolution layer from adversarial attacks. Moreover, as the size of the perturbation increases, the level of protection offered by IWMF also becomes correspondingly stronger.}
\label{fig_cnn}
\end{figure}

\noindent
\textbf{Security by adversarial purification.~}
Adversarial examples are generated by altering each pixel to mimic legitimate inputs. Typically, increasing the pixel differences enhances feature extraction, with the scale of these differences determined by the perturbation size, denoted as $\epsilon$. This characteristic is illustrated in Figure~\ref{fig_distribution}, where the difference values are computed by comparing each channel of the adversarial example to its corresponding source image. The blue bars in the figure indicate that most effective perturbations in the adversarial examples are equal to either $-\epsilon$ or $+\epsilon$, demonstrating maximum difference. However, following window mean processing, these perturbations are significantly reduced and tend to concentrate around zero, as shown by the orange bars. This effect is justified because the perturbations sum to zero, $(-\epsilon) + (+\epsilon)$, effectively canceling out the net change. Additionally, the difference at each pixel is distributed across nine pixels within the same window.

Figure~\ref{fig_cnn} demonstrates how IWMF enhances purifying the perturbations in the first convolution layer of deep learning models. As illustrated in Figure~\ref{fig_cnn}(b), the adversarial attack necessitates adding $\epsilon$ to five pixels in each window ($9\rightarrow10$ and $9\rightarrow8$) based on the convolution kernel to achieve optimal perturbing and reverse feature extraction results. The convolution outcomes of the adversarial example in Figure~\ref{fig_cnn}(b) exhibit significant differences ($45\rightarrow50$ and $45\rightarrow40$) in comparison to the original image displayed in Figure~\ref{fig_cnn}(a). However, after implementing IWMF to purify the windows, the convolution outcomes are highly preserved, with little discernible difference between the filtered image and the unmodified original image in Figure~\ref{fig_cnn}(a) ($45\rightarrow45$ and $45\rightarrow45$). Additionally, IWMF effectively mitigates adversarial attacks even when larger values of $\epsilon$ are employed. As illustrated in Figure~\ref{fig_cnn}(c), without defense, the outcomes of the first convolution layer exhibit more noticeable differences from the original image compared to Figure~\ref{fig_cnn}(b). However, after implementing IWMF, the extent of the difference is significantly reduced.

Furthermore, the purification process is further enhanced through the iterative processing feature of IWMF, where the averaged difference is propagated to the next window, and then subsequently further averaged.

\begin{figure}[t]
\setlength{\abovecaptionskip}{0.1cm}
\setlength{\belowcaptionskip}{0.1cm}
\centering
\includegraphics[width=0.9\linewidth]{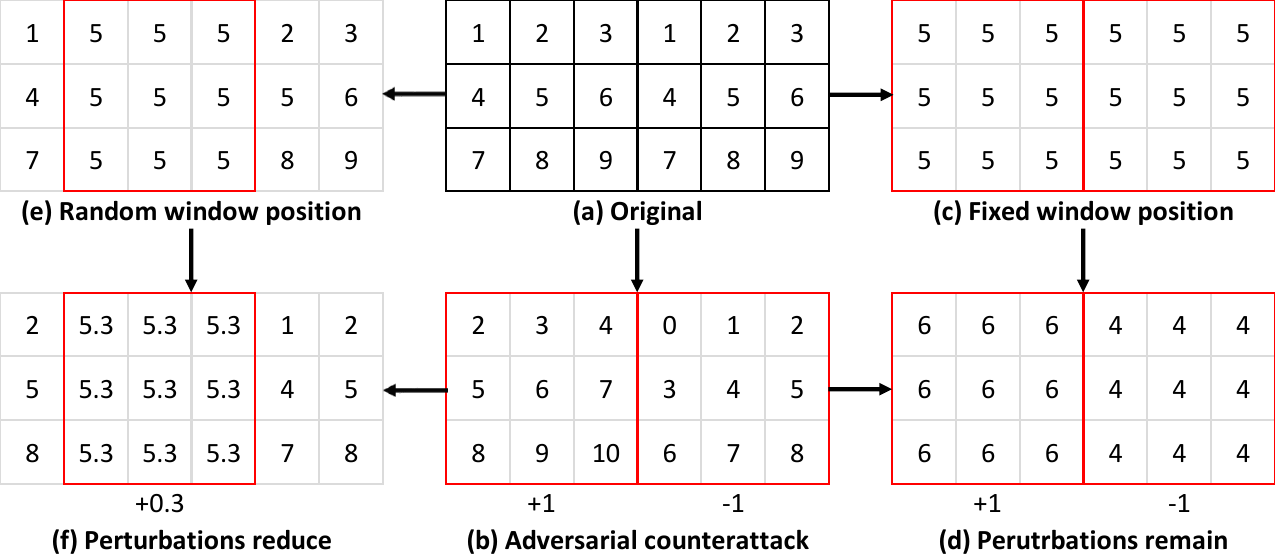}
\caption{Windows are randomly selected to resist adaptive attacks.}
\label{fig_random}
\end{figure}

\noindent
\textbf{Resistance against adaptive attacks.~}
IWMF's resistance against adaptive attacks is attributed to the random selection of windows. Figure~\ref{fig_random} depicts how if window positions and orders remain fixed (Figure~\ref{fig_random}(c)), adaptive attacks can learn to introduce perturbations to specific fixed windows in a specific order, allowing them to remain even after undergoing the purification process (Figure~\ref{fig_random}(d)). On the other hand, IWMF randomly selects windows (Figure~\ref{fig_random}(e)), thereby covering the ``well-designed'' perturbations introduced by adaptive attacks. Consequently, the perturbations are effectively covered (Figure~\ref{fig_random}(f)).

\noindent
\textbf{Generalization against various attacks.~}
The random selection of windows also plays a significant role in generalizing to various attacks. By randomizing the selection, IWMF can effectively purify perturbations regardless of any variation in the type, size, or location of the perturbations. If the window amount $\lambda$ is adequately large (\eg $(\lambda\times9>2)$), every pixel in the image will be changed at least once, indicating that the perturbation has been effectively averaged. However, larger values of $\lambda$ will undoubtedly intensify blurring effects on genuine images, so decrease overall accuracy. Therefore, the value of $\lambda$ should be carefully selected based on the specific requirements.

Additionally, by calculating the mean of both positive and negative difference values within each window, IWMF becomes more effective against attacks that use larger values of $\epsilon$, as evident in Figure~\ref{fig_cnn}(c), where the state-of-the-art defense mechanism fails to provide ample protection.

\subsection{Restoring IWMF-blurred Images by Diffusion Models}
\label{ddrm}
If the inputs are merely blurred by IWMF or other distortion methods, the resistance against adversarial examples can be ensured, but the accuracy of verifying genuine images will significantly decrease~\cite{wang2017adversary,XuW171,li2017adversarial,graese2016assessing,guo2018countering}. Hence, IWMF-Diff employs denoising diffusion models to restore blurred images while maintaining the accuracy of genuine inputs. To achieve this, additive Gaussian noise must be applied to the blurred images, as a compulsory condition discussed in this section. This Gaussian noise further covers adversarial perturbations. Note that individual diffusion-based methods (\eg the state-of-the-art approach DiffPure) without IWMF are futile against attacks in large perturbations sizes. This is because the settings of these methods (\eg slight Gaussian noise) are insufficient to cover large perturbations.

This section presents the reverse process of IWMF-Diff using DDRM. DDRM~\cite{kawar2022denoising} is a Gaussian-based denoising diffusion model capable of restoring images by reversing the diffusion process. It includes denoising, super-resolution, deblurring, inpainting, and colorization. In IWMF-Diff, DDRM's pretrained model on the CelebA dataset~\cite{Liu15} is directly utilized for face image restoration.

Given the pretrained model $p_\theta$, DDRM is defined as a Markov chain $x_T\rightarrow x_{T-1}\rightarrow\cdots\rightarrow x_1\rightarrow x_0$ conditioned on $y$ for any linear inverse task. Here,

\begin{equation}
p_\theta(x_{0:T}|y)=p_\theta(x_T|y)\prod_{t=0}^{T-1}p_\theta(x_t|x_{t+1},y),
\label{eq_ddrm_distribution}
\end{equation}
and $x_0$ is the final denoising output. 

In IWMF-Diff, the input of DDRM is the IWMF-blurred image $x^{IWMF}$. $x_0$ represents the expected restored image, while $y$ is $x^{IWMF}$ with the Gaussian noise in $\sigma_y$ satisfying:

\begin{equation}
y=\mathcal{N}(x^{IWMF},\sigma_y^2).
\label{eq_ddrm_y}
\end{equation}

The denoising strategy is most effective for adversarial purification as DDRM adds Gaussian noise to adversarial examples to cover perturbations. For denoising, Equation~\ref{eq_ddrm_distribution} can be formulated as follows:

\begin{equation}
p_\theta(x_T|y)=\mathcal{N}(y,\sigma_T^2-\sigma_y^2),
\end{equation}
\begin{equation}
\begin{aligned}
&p_\theta(x_t|x_{t+1},y)\\
&=\left\{\begin{array}{rcl}
\mathcal{N}(x_{t+1}+\sqrt{1-\eta^2}\sigma_t\frac{y-x_{t+1}}{\sigma_y},\eta^2\sigma_t^2)&\ s.t.\ \sigma_t<\sigma_y\\
\mathcal{N}((1-\eta_b)x_{t+1}+\eta_b y,\sigma_t^2-\eta_b^2\sigma_y^2)&\ s.t.\ \sigma_t\geq\sigma_y,
\end{array}
\right.
\end{aligned}
\label{eq_ddrm_reverse}
\end{equation}
where $\eta_*\in (0,1]$ is a hyperparameter that controls the variance of the transitions. When $\eta_b=1$, the maximum timestep of the reverse process is conditioned on $\sigma_y$. When $\eta=1$, the reverse process does not refer to any information of $y$. In IWMF-Diff, $\eta=0.85$ and $\eta_b=1$ as recommended by~\cite{kawar2022denoising} for optimal results. Therefore, Equation~\ref{eq_ddrm_reverse} can be further simplified as:

\begin{equation}
\begin{aligned}
&p_\theta(x_t|x_{t+1},y)=\\
&\left\{\begin{array}{cl}
\mathcal{N}(x_{t+1}+\frac{\sqrt{1-0.85^2}\sigma_t(y-x_{t+1})}{\sigma_y},0.85^2\sigma_t^2)&s.t.\ \sigma_t<\sigma_y\\
\mathcal{N}(y,\sigma_t^2-\sigma_y^2)&s.t.\ \sigma_t\geq\sigma_y.
\end{array}
\right.
\end{aligned}
\label{eq_ddrm_reverse_final}
\end{equation}

Replacing $y$ by Equation~\ref{eq_ddrm_y}, the image restoration process for the IWMF-blurred input $x^{IWMF}$ can be introduced as:

\begin{equation}
p_\theta(x_T|x^{IWMF})=\mathcal{N}(x^{IWMF},\sigma_T^2),
\end{equation}
\begin{equation}
\begin{aligned}
&p_\theta(x_t|x_{t+1},x^{IWMF})=\\
&\left\{\begin{array}{cl}
\mathcal{N}\left(x_{t+1}+\frac{\sqrt{1-0.85^2}\sigma_t(x^{IWMF}-x_{t+1})}{\sigma_y},\sigma_t^2\right)&s.t.\ \sigma_t<\sigma_y\\
\mathcal{N}(x^{IWMF},\sigma_t^2)&s.t.\ \sigma_t\geq\sigma_y.
\end{array}
\right.
\end{aligned}
\label{eq_ddrm_reverse_iwmf}
\end{equation}

As seen in Equation~\ref{eq_ddrm_reverse_iwmf}, the diffusion reverse process is conditioned at the maximum timestep $\sigma_y$ when $\sigma_y>0$:

\begin{equation}
p_\theta(x_T|x^{IWMF})=\mathcal{N}(x^{IWMF},\sigma_T^2)\ s.t.\ \sigma_T=\sigma_y.
\label{eq_ddrm_max}
\end{equation}

Equation \ref{eq_ddrm_max} indicates that effective image restoration always begins from $\sigma_y$. Additionally, DiffPure~\cite{nie2022DiffPure} has shown that Gaussian noise is effective in covering perturbations from adversarial examples. In other words, $x^{IWMF}$ in $\sigma_y$ can be seen as further purified for adversarial defense. To summarize, IWMF-Diff first blurs images using IWMF in $\lambda$ and then inputs them into DDRM. DDRM further replaces the adversarial perturbations with the Gaussian noise in $\sigma_y$.
Finally, DDRM regards Gaussian-blurred images as $y$ in Equation~\ref{eq_ddrm_reverse_final} and restores them using Equation~\ref{eq_ddrm_reverse_final} from $\sigma_T=\sigma_y$. Algorithm~\ref{alg_iwmf_diff} presents the IWMF-Diff process. Note that the proof of applicable image restoration using DDRM is provided by \cite{kawar2022denoising}, while the proof of feasible adversarial purification using Gaussian noise can be found in \cite{nie2022DiffPure}.

\begin{algorithm}[t]
\caption{IWMF-Diff}
\label{alg_iwmf_diff}
\begin{algorithmic}[1]
\Require{Image X, window amount $\lambda$, window size s, Gaussian standard deviation $\sigma_y>0$, pretrained diffusion model $\theta$} 
\Ensure{Purified image $x_0$}
\State \(x^{IWMF} \leftarrow IWMF(X,\lambda,s)\) \algorithmiccomment{refer to Algorithm~\ref{alg_iwmf}}
\State \(y \leftarrow \mathcal{N}(x^{IWMF},\sigma_y^2)\)
\State \(x_T \leftarrow y\)
\State \(\sigma_T \leftarrow \sigma_y\)
\For{t in [T-1:0]}
\State \(\begin{aligned}
&p_\theta(x_t|x_{t+1},y)\\
&=\mathcal{N}(x_{t+1}+\sqrt{1-0.85^2}\sigma_t\frac{y-x_{t+1}}{\sigma_y},0.85^2\sigma_t^2)
\end{aligned}\)
\EndFor
\State \Return{$x_0$}
\end{algorithmic}
\end{algorithm}

There is a special case in Equation~\ref{eq_ddrm_reverse_iwmf}, which occurs when $\sigma_y=0$. In this case:

\begin{equation}
x_0=\mathcal{N}(x^{IWMF},\sigma_0^2).
\end{equation}

According to the configuration of the pretrained model on the CelebA dataset, $\sigma_0=0.0001$. However, since $\sigma_0$ is too small, neither purification nor restoration using DDRM is feasible. Therefore, the addition of Gaussian noise to the IWMF-blurred images is essential for image restoration, not just for better purification.

\section{Experimental Settings}
% \subsection{Experimental settings}
\subsection{Deep Learning Models for Face Authentication}
Major evaluations and analysis are conducted on InsightFace~\cite{deng2019arcface} for several reasons: \one\ InsightFace is one of the most widely used and best-performing deep learning models for face authentication, and is the backbone of many commercial APIs (\eg Amazon Rekognition). Protecting InsightFace leads to improved performance in existing systems. \two\ InsightFace has been shown to be vulnerable to adversarial attacks. \three\ For fair comparison, benchmark defenses~\cite{ren2022perturbation,zhou2020manifold,nie2022DiffPure} employ InsightFace as the backbone and conduct experiments on it. However, extra evaluations are conducted on FaceNet \cite{schroff2015facenet} to investigate the applicability of IWMF-Diff to different models (\ie authentication systems). Both models use 512-dimensional facial features after feature extraction.

\subsection{Datasets}
DDRM is trained on the CelebA dataset~\cite{Liu15}. All evaluations are conducted on the Labeled Faces in the Wild (LFW) dataset~\cite{Hua08}. For each evaluation, 500 adversarial examples are generated for 50 subjects. Note that the proposed IWMF-Diff does not require any further training, so an excessive number of samples is unnecessary for the experiments.

\subsection{Adversarial Attacks to Defend}
To assess the effectiveness of the proposed defenses, five benchmark gradient-based white-box attacks (FGSM~\cite{goodfellow2014explaining}, PGD~\cite{madry2018towards}, CW~\cite{carlini2017towards}, APGD~\cite{croce2020reliable,croce2021robustbench,pintor2022indicators}, and SGADV~\cite{wang2021similarity} (face-specific)), one gradient-based black-box attack (BIM)~\cite{kurakin2018adversarial}, one query-based black-box attack (Square attack)~\cite{andriushchenko2020square}, and three facial-landmark-based black-box attacks (TI-FGSM~\cite{dong2019evading}, DI$^2$-FGSM~\cite{xie2019improving}, and LGC~\cite{yang2020robfr} (face-specific)) are selected. The attack settings are based on the recommendations in their respective papers and are listed in Table~\ref{tab_attack}.

\begin{table}[t]
\setlength{\abovecaptionskip}{0.1cm}
\setlength{\belowcaptionskip}{0.1cm}
\caption{Settings of the adversarial attacks}
\label{tab_attack}
\centering
\setlength{\tabcolsep}{1mm}{\begin{tabular}{cc}
\hline
Technique&Settings\\
\hline
\hline
FGSM~\cite{goodfellow2014explaining}&$\epsilon=0.03$\\
PGD~\cite{madry2018towards}&$\epsilon=0.03$, $\alpha=0.001$, $t_{max}=40$\\
CW~\cite{carlini2017towards}&\begin{tabular}{c}$\epsilon=0.03$, $\alpha=0.001$, $t_{max}=1,000$,\\binary search iterations = 20\end{tabular}\\
SGADV~\cite{wang2021similarity}&$\epsilon=0.03$, $\alpha=0.001$, $t_{max}=1,000$, $\tau_{conv}=0.0001$\\
APGD~\cite{croce2020reliable}&$\epsilon=0.03$, $t_{max}=40$\\
APGD-EOT~\cite{lee2023robust}&$\epsilon=0.03$, $t_{max}=40$, EOT iteration = 20\\
\hline
BIM~\cite{kurakin2018adversarial}&$\epsilon=4/255$, $\alpha=0.001$, $t_{max}=20$\\
TI-FGSM~\cite{dong2019evading}&$\epsilon=4/255$, $\alpha=0.001$, $t_{max}=20$, $m=4$, $\mu=1$\\
DI$^2$-FGSM~\cite{xie2019improving}&$\epsilon=4/255$, $\alpha=0.001$, $t_{max}=20$, $m=4$, $\mu=1$\\
LGC~\cite{yang2020robfr}&$\epsilon=4/255$, $\alpha=0.001$, $t_{max}=20$, $m=4$, $\mu=1$\\
Square~\cite{andriushchenko2020square}&$\epsilon=0.03$, $t_{max}=20,000$\\
\hline
\end{tabular}}
\end{table}

\subsection{Benchmark Defenses}
The proposed methods are compared with two latest auto-encoder-based methods, \ie A-VAE~\cite{zhou2020manifold} and PIN~\cite{ren2022perturbation}, along with the state-of-the-art Gaussian-diffusion-based adversarial purification method called DiffPure~\cite{nie2022DiffPure}. All these methods are claimed to be applicable for face authentication and have demonstrated their superiority over other defenses~\cite{goodfellow2014explaining,xie2019feature,jia2019comdefend,guo2018countering,song2018pixeldefend,meng2017magnet,zhong2019adversarial,liao2018defense,vahdat2020nvae,karras2020analyzing,chai2021ensembling,richardson2021encoding}. 
Please refer more details about benchmark defenses in Appendix~A.
It should be noted that A-VAE does not release its code or pre-trained model. Therefore, we have used the experimental results quoted in their paper and followed the same protocols for conducting our experiments to ensure a fair comparison. Furthermore, we conduct an ablation study comparing the proposed non-deep-learning IWMF with six traditional non-deep-learning defenses.

\subsection{Adaptive Attacks}
Breaching systems protected by defense modules is the primary objective of adaptive attacks, making the design of effective defense strategies particularly challenging. To evaluate the effectiveness of our proposed defense mechanisms against adaptive attacks, we adopted several approaches. First, considering the randomization strategies integrated into the evaluated defense algorithms, we employed Expectation over Transformation (EOT) \cite{athalye2018obfuscated,pintor2022indicators,croce2022evaluating,lee2023robust} to attack randomized defenses by optimizing the expectation of the randomness (see Table~\ref{tab_attack}). Second, we reformulated SGADV \cite{wang2021similarity} as an algorithm-specific adaptive attack against the defense methods, including PIN \cite{ren2022perturbation}, DiffPure \cite{nie2022DiffPure}, the proposed IWMF, and IWMF-Diff. In the strong white-box setting, adaptive attacks have complete knowledge of the deep learning model, database, and defense mechanisms \cite{carlini2017adversarial}. The implementations of these adaptive attacks are detailed in Algorithms~\ref{alg_ca_pin} and \ref{alg_ca_diff}. Finally, we assessed the defenses by applying the reformulated adaptive attack algorithm, omitting the randomization strategies in DiffPure, IWMF, and IWMF-Diff to evaluate whether randomization enhances resistance against adaptive attacks.

\subsection{Evaluation Metrics}
We involve the following metrics to evaluate the proposed IWMF-Diff framework.

\begin{algorithm}[t]
    \caption{Adaptive SGADV attacking PIN}
    \label{alg_ca_pin}
    \begin{algorithmic}[1]
     \Require{Source image $X^S$, target image $X^T$, feature extractor $PIN(\cdot)$, perturbation size $\epsilon$, step size $\alpha$, maximum steps $t_{max}$} 
     \Ensure{Adversarial example $X^{adv}$}
     \State \(\delta^0\sim U(-\epsilon,\epsilon)\)
     \State \(X^0 \leftarrow X^S+\delta^0\)
     \Repeat
      \State \(J_{SG}(X^{t},X^T)=||PIN(X^{t})-PIN(X^T)||\)
      \State \(X^{t+1}=Clip_{X^S,\epsilon}\{X^{t}+\alpha \cdot sign(\nabla_{X^{t}}J_{SG})\}\)
      \Until{convergence~\cite{wang2021similarity} or \(t=t_{max}\)}
     \State \Return{\(X^{adv} \leftarrow X^{t_{stop}}\)}
    \end{algorithmic}
\end{algorithm}
\begin{algorithm}[t]
    \caption{Adaptive SGADV attacking DiffPure, IWMF, and IWMF-Diff}
    \label{alg_ca_diff}
    \begin{algorithmic}[1]
     \Require{Source image $X^S$, target image $X^T$, feature extractor $f(\cdot)$, perturbation size $\epsilon$, step size $\alpha$, maximum steps $t_{max}$, window amount $\lambda$, window size $s$, Gaussian standard deviation $\sigma_y$} 
     \Ensure{Adversarial example $X^{adv}$}
     \State \(\delta^0\sim U(-\epsilon,\epsilon)\)
     \State \(X^0 \leftarrow X^S+\delta^0\)
     \Repeat
      \State \(X^{tmp} = IWMF(X^{t}|\lambda,s)\)\algorithmiccomment{it is DiffPure when \(\lambda=0\)}
      \State \(X^{tmp} = Diff(X^{tmp}|\sigma_y)\)\algorithmiccomment{it is IWMF without this step}
      \State \(J_{SG}(X^{tmp},X^T)=||f(X^{tmp})-f(X^T)||\)
      \State \revise{\(X^{t+1}=Clip_{X^S,\epsilon}\{X^{t}+\alpha \cdot sign(\nabla_{X^{tmp}}J_{SG})\}\)}
      \Until{convergence~\cite{wang2021similarity} or \(t=t_{max}\)}
     \State \Return{\(X^{adv} \leftarrow X^{t_{stop}}\)}
    \end{algorithmic}
\end{algorithm}

\noindent \textbf{False reject rate (FRR)}~\cite{Jain2011Introduction} refers to the probability that the authentication system falsely rejects a genuine image. In particular, $FRR_{genuine}$ and $FRR_{attack}$ (\eg $FRR_{FGSM}$) reflect the accuracy of correctly classifying genuine images and adversarial examples as their true identities, respectively. 
A smaller value is preferred for both types of FRR. It is important to note that the true accept rate (TAR) is simply calculated as $1 - FRR$ and is used in generating the Receiver Operating Characteristic (ROC) curves.

\noindent\textbf{False accept rate (FAR)}~\cite{Jain2011Introduction} refers to the probability of the authentication system falsely accepting an imposter image or adversarial example. A smaller value is preferred for FAR. In particular, adversarial examples are considered as imposter images, and hence, $FAR_{attack}$ denotes the attack success rate specifically for adversarial examples.

\noindent \textbf{Equal error rate (EER)}~\cite{Jain2011Introduction} is a metric used to evaluate the effectiveness of an authentication system. It indicates the point on the ROC curve where the FAR equals the FRR. It is generally preferred that the EER of an authentication system is smaller as it indicates better performance.

\noindent \textbf{Area under the ROC curve (AUC)}~\cite{Jain2011Introduction} is a measure of the overall performance of an authentication system across all possible classification thresholds. It quantifies the entire two-dimensional area beneath the ROC curve, which represents the TAR against the FAR. A higher AUC implies better performance, making it a preferred evaluation metric.

\noindent \textbf{Cosine similarity} \cite{deng2019arcface} 
is a qualitative measure that evaluates the similarity between two images in the feature space. In the case of genuine images and enrolled images, the higher the Cosine Similarity, the better the closeness between them. On the other hand, when dealing with adversarial examples and target images, a lower cosine similarity is preferred as it implies lesser likelihood of the adversarial example being successful to fool the target system.
% \begin{equation}
% 	similarity=0.5-0.5\times cos[f(X_1),f(X_2)] \in [0,1].
% \end{equation}
% where $f(\cdot)$ is the feature extractor and $X_*$ is the image.

\subsection{System Settings}
\label{settings}
In the context of our experiments, the term ``one system'' refers to a deep learning model that is integrated with a defense mechanism, if any, and works with a particular database. As such, when conducting experiments on this ``one system'', we maintain the same settings for both the authentication and defense operations, applying them uniformly across all attacks. As per our experiment, the authentication accuracy and security are influenced by four parameters, namely threshold ($\tau$), window amount ($\lambda$), window size ($s$), and Gaussian standard deviation ($\sigma_y$). Specifically, we have listed the settings for each case in Table~\ref{tab_settings}. It is important to note that these settings ensure optimal performance against SGADV.

\begin{table*}[t]
 \centering
 \begin{threeparttable}
 	\setlength{\abovecaptionskip}{0.1cm}
	\setlength{\belowcaptionskip}{0.1cm}
  \caption{System setting list}
  \label{tab_settings}
  \setlength{\tabcolsep}{5.7mm}{
   \begin{tabular}{ccccccc}
    \hline
    Deep model&Defense&Condition&$\lambda$&$\sigma_y$&$\tau$&$s$\\
    \hline
    \hline
    \multirow{5}{*}{InsightFace}&No defense&\(FRR_{genuine}=FAR_{imposter}\)&N/A&N/A&0.6131&N/A\\
    \Xcline{2-7}{0.4pt}
    &PIN~\cite{ren2022perturbation}&\multirow{4}{*}{\(FRR_{genuine}=FAR_{SGADV}\)}&N/A&N/A&0.5890&N/A\\
    &DiffPure~\cite{nie2022DiffPure}&&N/A&0.15&0.7119&N/A\\
    &IWMF (ours)&&0.40&N/A&0.6611&3px\\
    &IWMF-Diff (ours)&&0.25&0.15&0.6351&3px\\
    \hline
    \multirow{5}{*}{FaceNet}&No defense&\(FRR_{genuine}=FAR_{imposter}\)&N/A&N/A&0.7056&N/A\\
    \Xcline{2-7}{0.4pt}
    &PIN~\cite{ren2022perturbation}&\multirow{4}{*}{\(FRR_{genuine}=FAR_{SGADV}\)}&N/A&N/A&0.5890&N/A\\
    &DiffPure~\cite{nie2022DiffPure}&&N/A&0.15&0.7407&N/A\\
    &IWMF (ours)&&0.85&N/A&0.7052&3px\\
    &IWMF-Diff (ours)&&0.20&0.15&0.7117&3px\\
    \hline
   \end{tabular}}
 \end{threeparttable}
\end{table*}

\begin{figure}[b]
\setlength{\abovecaptionskip}{0.1cm}
\setlength{\belowcaptionskip}{0.1cm}
\centering
\includegraphics[width=0.7\linewidth]{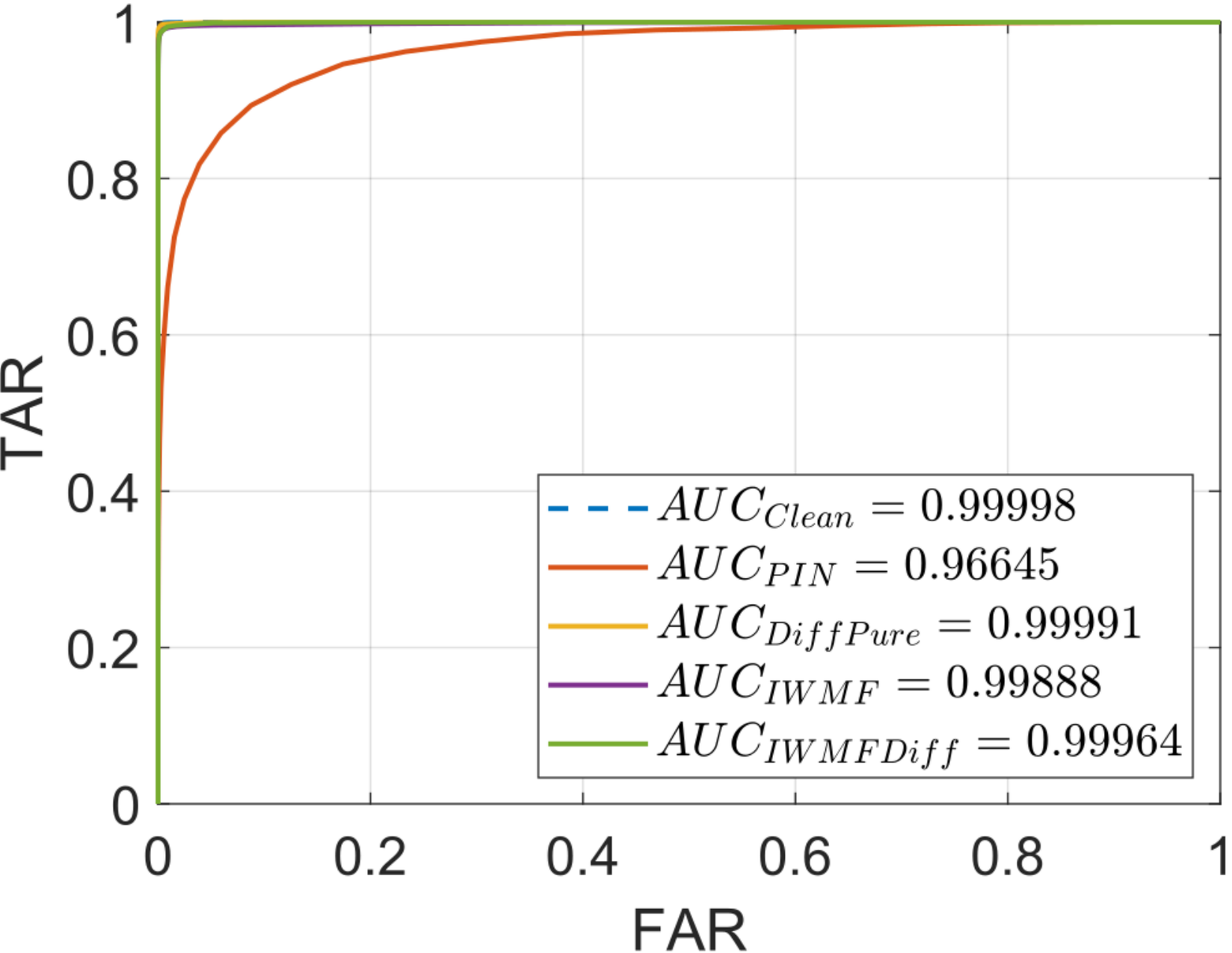}
\caption{
ROC curves for benign (non-adversarial) images using various defensive techniques applied to InsightFace. The results indicate that all defenses except PIN preserve accuracy for benign images effectively.}
\label{fig_roc_insightface}
\end{figure}

\section{Experiment Results}
\label{df_performance}

\subsection{Requirements of Ideal Adversarial Defenses}
\label{df_performance_4}

\noindent \textbf{Authentication accuracy against genuine images.~} 
The AUC scores presented in Figure~\ref{fig_roc_insightface} reveal that the performance of the proposed IWMF and IWMF-Diff against genuine images is highly effective as compared to the original system without defense modules ($AUC_{orig}$), with a maximum decrease in AUC scores of 0.0035. However, the auto-encoder-based PIN is not suitable for authenticating genuine images, as it exhibits a maximum decrease in AUC scores of 0.0335. The outcomes of $FRR_{genuine}$, as listed in Table~\ref{tab_df_if_sgadv_w}, demonstrate that the proposed IWMF-Diff surpasses other benchmark defenses in the same protocol deemed best performing against SGADV. The error rate exhibits a drop from 5\% to 3.22\%, which is better than the state-of-the-art defense. We would like to mention that we could not produce ROC curves for A-VAE as it has not made its code or pre-trained model publicly available.

\begin{table*}[!t]
\setlength{\abovecaptionskip}{0.1cm}
\setlength{\belowcaptionskip}{0.1cm}
\centering
\begin{threeparttable}
\caption{Error (\%) of falsely rejecting genuine images (FRR) and accepting white-box adversarial examples (FAR) in InsightFace}
\label{tab_df_if_sgadv_w}
\setlength{\tabcolsep}{5mm}{
\begin{tabular}{c|c|ccccc}
\hline
Defense&$FRR_{genuine}$&$FAR_{SGADV}$&$FAR_{FGSM}$&$FAR_{PGD}$&$FAR_{CW}$&$FAR_{APGD}$\\
\hline
\hline
InsightFace&0.28&100.0&100.0&100.0&100.0&100.0\\
\Xcline{1-7}{0.8pt}
A-VAE$^1$~\cite{zhou2020manifold}&5.90&-&23.7&36.1&-&-\\
PIN~\cite{ren2022perturbation}&17.60&16.4&18.4&15.4&13.8&24.2\\
DiffPure~\cite{nie2022DiffPure}&5.00&5.0&32.8&\textbf{0.4}&\textbf{0.0}&17.4\\
IWMF (ours)&6.36&6.2&16.2&1.0&\textbf{0.0}&9.2\\
IWMF-Diff (ours)&\textbf{3.22}&\textbf{3.2}&\textbf{15.6}&0.8&0.2&\textbf{6.6}\\
\hline
\hline
IWMF-Diff (fair)$^2$&5.00&1.0&11.2&0.0&0.0&-\\
\hline
\end{tabular}}
\begin{tablenotes}
\footnotesize
\item $^{1}$A-VAE has neither released the code nor pretrained model. The numbers are quoted from its paper, so do not include results against CW and SGADV.
\item $^{2}$``fair'' represents that this row is for the fair comparison by adjusting the threshold to make one of the results same as the baseline.
\end{tablenotes}
\end{threeparttable}
\end{table*}

\begin{table*}[!t]
\setlength{\abovecaptionskip}{0.1cm}
\setlength{\belowcaptionskip}{0.1cm}
\centering
\begin{threeparttable}
\caption{Error (\%) of falsely rejecting genuine images (FRR) and accepting black-box adversarial examples (FAR) in InsightFace}
\label{tab_df_if_sgadv_b}
\setlength{\tabcolsep}{4.2mm}{
\begin{tabular}{c|c|ccccc}
\hline
Defense&$FRR_{genuine}$&$FAR_{DI^2-FGSM}$&$FAR_{TI-FGSM}$&$FAR_{LGC}$&$FAR_{BIM}$&$FAR_{Square}$\\
\hline
\hline
InsightFace&0.28&95.00&93.17&93.73&91.50&100\\
\Xcline{1-7}{0.8pt}
DiffPure~\cite{nie2022DiffPure}&5.00&41.67&37.27&36.77&29.67&20.4\\
IWMF (ours)&6.36&\textbf{4.63}&\textbf{6.27}&\textbf{3.37}&\textbf{1.17}&28.8\\
IWMF-Diff (ours)&\textbf{3.22}&28.53&33.00&23.97&10.87&\textbf{19.8}\\
\hline
\end{tabular}}
\end{threeparttable}
\end{table*}

\noindent \textbf{Defense against white-box adversarial attacks.~} 
As demonstrated in Table~\ref{tab_df_if_sgadv_w}, the proposed IWMF and IWMF-Diff defense mechanisms significantly enhance the security of previously vulnerable deep learning models by substantially reducing the attack success rates, measured as FARs, in comparison to baseline models. The IWMF-Diff defense, in particular, outperforms benchmark defenses, showing superior efficacy. Although IWMF-Diff exceeds the performance of both PIN and A-VAE, it is worth noting that its $FAR_{PGD}$ and $FAR_{CW}$ values are slightly higher than those of DiffPure, primarily due to differing threshold settings (refer to Table~\ref{tab_settings}). To ensure a fair comparison between IWMF-Diff and DiffPure, we conducted an analysis by maintaining equal $FRR_{genuine}$ values for both methods. The results indicate that IWMF-Diff offers greater security against all white-box attacks than DiffPure. Furthermore, the IWMF defense mechanism delivers performance comparable to the state-of-the-art defense DiffPure. The FAR values of IWMF against PGD and CW attacks are comparable to those of DiffPure. While the $FAR_{SGADV}$ and $FRR_{genuine}$ values of IWMF are higher than those of DiffPure, the $FAR_{FGSM}$ and $FAR_{APGD}$ values of IWMF are lower.

\noindent \textbf{Defense against black-box adversarial attacks.~} 
The results listed in Table~\ref{tab_df_if_sgadv_b} suggest that the robustness of IWMF and IWMF-Diff against black-box attacks is dramatically enhanced and outperforms the state-of-the-art defense DiffPure by a considerable margin. Among the two, IWMF is deemed more suitable as the Gaussian-based diffusion employed by IWMF-Diff only marginally helps in concealing perturbations resulting from black-box adversarial examples. A more detailed discussion on this observation is presented in Section \ref{df_ablation_blur_diff}.

\noindent \textbf{Robustness of classifying adversarial examples.~} 
The results presented in Table~\ref{tab_rb_if_sgadv_w} indicate that IWMF-Diff significantly enhances the robustness of the deep learning model in classifying adversarial examples as their true labels, where other defense mechanisms are generally not viable. We would like to note that the $FRR_{FGSM}$ of IWMF-Diff is slightly higher than the original system without defense as it uses a higher classification threshold to defend attacks. This suggests that the purification of adversarial examples can enhance the model's resistance against such attacks. However, since the images have been perturbed, blurred, and then restored, the accuracy is typically lower than that of non-adversarial genuine images.

\begin{table*}[!t]
\setlength{\abovecaptionskip}{0.1cm}
\setlength{\belowcaptionskip}{0.1cm}
\centering
\begin{threeparttable}
\caption{Error (\%) of falsely rejecting adversarial examples (FRR) as their true identities in InsightFace}
\label{tab_rb_if_sgadv_w}
\setlength{\tabcolsep}{5mm}{
\begin{tabular}{c|c|ccccc}
\hline
Defense&$FRR_{genuine}$&$FRR_{SGADV}$&$FRR_{FGSM}$&$FRR_{PGD}$&$FRR_{CW}$&$FRR_{APGD}$\\
\hline
\hline
InsightFace&0.28&98.30&6.34&51.92&42.12&95.20\\
\Xcline{1-7}{0.8pt}
PIN~\cite{ren2022perturbation}&17.60&18.88&16.46&17.86&17.60&21.24\\
DiffPure~\cite{nie2022DiffPure}&5.00&20.08&28.66&13.00&7.68&30.08\\
IWMF (ours)&6.36&25.50&19.58&17.38&13.08&26.72\\
IWMF-Diff(ours)&\textbf{3.22}&\textbf{12.06}&\textbf{8.28}&\textbf{9.22}&\textbf{6.18}&\textbf{9.22}\\
\hline
\end{tabular}}
\end{threeparttable}
\end{table*}

\noindent \textbf{Generalization against various attack algorithms.~}
The results presented in Tables~\ref{tab_df_if_sgadv_w} to \ref{tab_rb_if_sgadv_w} demonstrate that the robustness of IWMF and IWMF-Diff is considerably enhanced, indicating their outstanding generalization capability against various types of attacks.

\begin{figure}[t]
\centering
\includegraphics[width=0.7\linewidth]{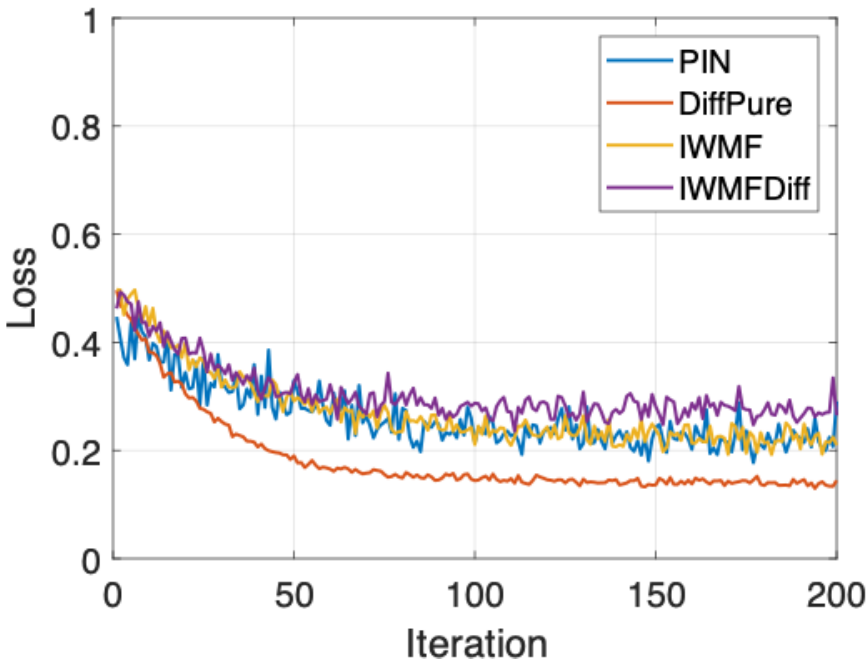}
\caption{
The IWMF-Diff exhibits the least amount of convergence, yet the state-of-the-art defense, DiffPure, shows weakest performance against adaptive attacks.
}
\label{fig_convergence}
\end{figure}

\noindent \textbf{Resistance against adaptive attacks.~}
Defending against adaptive attacks poses a significant challenge for adversarial defenses. We evaluated the reliability of the proposed defenses and compared them with benchmark defenses by deploying the APGD-EOT attack \cite{athalye2018obfuscated,pintor2022indicators,croce2022evaluating,lee2023robust} and our specially designed algorithm-specific adaptive attack (Algorithms \ref{alg_ca_pin} and \ref{alg_ca_diff}). The findings, presented in Table~\ref{tab_counterattack_if}, reveal several key insights. \one~Although EOT was used to attack randomized defenses by optimizing the expectation of randomness, its FARs are not significantly higher than those observed without EOT ($FAR_{SGADV}$). This suggests that the randomization strategy integrated into these defenses effectively enhances their resistance to adaptive attacks. \two~Our newly designed adaptive attack demonstrates considerable aggression in breaching defenses. Nevertheless, the proposed IWMF and IWMF-Diff defenses exhibit better efficacy. Figure~\ref{fig_convergence} shows that the loss associated with IWMF-Diff is most converged, whereas the state-of-the-art defense DiffPure shows the weakest performance against adaptive attacks. \three~To assess the necessity of the randomization strategy, we conducted adaptive attacks with the randomization in defenses disabled. The resulting increases in FARs (denoted by $^*$) underscore the importance of the randomization strategy in the proposed IWMF and IWMF-Diff, highlighting its effectiveness.

\begin{table}[t]
\setlength{\abovecaptionskip}{0.1cm}
\setlength{\belowcaptionskip}{0.1cm}
\centering
\begin{threeparttable}
\caption{Resistance (\%) to adaptive attacks for InsightFace}
\label{tab_counterattack_if}
\setlength{\tabcolsep}{1.2mm}{
\begin{tabular}{c|c|cc}
\hline
Defense&$FAR_{SGADV}$&$FAR_{APGD-EOT}$&$FAR_{adaptive}$\\
\hline
\hline
InsightFace&100.0&100.0&N/A\\
\Xcline{1-4}{0.8pt}
PIN~\cite{ren2022perturbation}&16.4&20.4&87.6\\
DiffPure~\cite{nie2022DiffPure}&5.0&17.6&99.4/98.8$^*$\\
IWMF (ours)&6.2&7.6&80.4/98.8$^*$\\
IWMF-Diff (ours)&3.2&\textbf{5.0}&\textbf{77.4}/92.0$^*$\\
\hline
\end{tabular}}
\begin{tablenotes}
\footnotesize
\item $^{*}$ denotes that the defense does not introduce randomization.
\end{tablenotes}
\end{threeparttable}
\end{table}

\subsection{Computational Complexity}
\label{df_time}
We evaluated the time efficiency of adversarial defenses but we believe that the requirement for this metric is determined by the specific use cases. Table 6 shows comparable time costs for strategies of the same type (e.g., blurring or diffusion). However, diffusion-based denoising takes significantly more time compared to blurring, making it difficult to apply diffusion-based adversarial defenses in real-time tasks. Referring to Table 10, to achieve better efficiency in real-time tasks, it is recommended to use IWMF without diffusion. However, for the highest level of security, IWMF-Diff provides superior performance.

\begin{table}[!t]
\setlength{\abovecaptionskip}{0.1cm}
\setlength{\belowcaptionskip}{0.1cm}
\centering
\begin{threeparttable}
\caption{Time cost (s) of processing $112\times112$ images @ $\lambda=0.25$, $\sigma_y=0.15$}
\label{tab_time}
\setlength{\tabcolsep}{5mm}{
\begin{tabular}{ccc}
\hline
Strategy&Single&500\\
\hline
Gaussian&0.01&0.06\\
IWMF (ours)&0.36&0.37\\
noisless diffusion (DDPM~\cite{ho2020denoising})&3.41&458.09\\
IWMF+diffusion&3.80&458.50\\
Gaussian+diffusion (DiffPure~\cite{nie2022DiffPure})&3.41&458.10\\
IWMF+Gaussian+diffusion (ours)&3.79&458.51\\
\hline
\end{tabular}}
\end{threeparttable}
\end{table}

\subsection{Generalization to Other Deep Learning Models}
\label{df_facenet}
The proposed IWMF and IWMF-Diff are designed as pre-processing modules before authentication. In section~\ref{df_performance_4}, these two methods were shown to be superior in protecting InsightFace. In this section, we further evaluate the ability of IWMF and IWMF-Diff to generalize against other deep learning models, such as FaceNet.

\begin{figure}[t]
\setlength{\abovecaptionskip}{0.1cm}
\setlength{\belowcaptionskip}{0.1cm}
\centering
\includegraphics[width=0.7\linewidth]{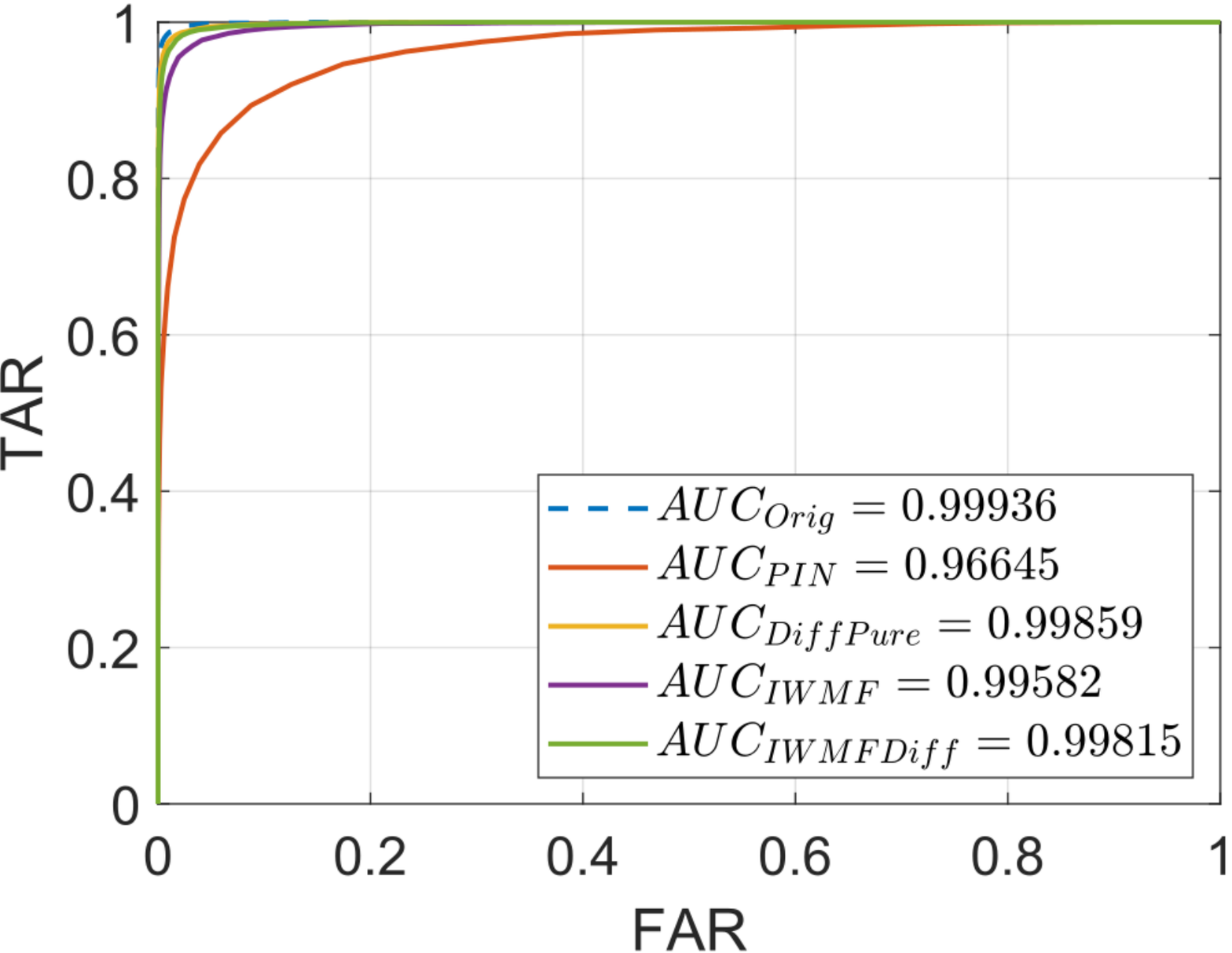}
\caption{
The ROC curves display the performance of various defense methods in protecting FaceNet against benign (non-adversarial) images. Among the defenses, only PIN is ineffective in preserving accuracy for benign images.
}
\label{fig_roc_facenet}
\end{figure}

Figure~\ref{fig_roc_facenet} and Tables~\ref{tab_df_fn_sgadv_w} to \ref{tab_counterattack_fn} illustrate that IWMF and IWMF-Diff outperform benchmark defenses in defending FaceNet. These results indicate that the proposed methods are applicable to all existing face authentication systems. Furthermore, Table~\ref{tab_df_fn_sgadv_b} shows that the state-of-the-art defense DiffPure is ineffective against black-box attacks, as the FARs are nearly identical to those without any defense.

\begin{table*}[t]
\setlength{\abovecaptionskip}{0.1cm}
\setlength{\belowcaptionskip}{0.1cm}
\centering
\begin{threeparttable}
\caption{Error (\%) of falsely rejecting genuine images (FRR) and accepting white-box adversarial examples (FAR) in FaceNet}
\label{tab_df_fn_sgadv_w}
\setlength{\tabcolsep}{5mm}{
\begin{tabular}{c|c|ccccc}
\hline
Defense&$FRR_{genuine}$&$FAR_{SGADV}$&$FAR_{FGSM}$&$FAR_{PGD}$&$FAR_{CW}$&$FAR_{APGD}$\\
\hline
\hline
FaceNet&1.20&100.0&91.2&100.0&100.0&100.0\\
\Xcline{1-7}{0.8pt}
PIN~\cite{ren2022perturbation}&17.60&15.0&\textbf{17.4}&12.0&11.8&16.4\\
DiffPure~\cite{nie2022DiffPure}&5.06&5.0&28.6&\textbf{1.4}&\textbf{0.6}&12.8\\
IWMF (ours)&6.38&6.4&20.2&3.6&2.2&9.6\\
IWMF-Diff (ours)&\textbf{3.80}&\textbf{3.8}&18.6&1.6&0.8&\textbf{8.6}\\
\hline
\hline
IWMF-Diff (fair)&5.04&3&13.4&0.8&0.4&-\\
\hline
\end{tabular}}
\end{threeparttable}
\end{table*}

\begin{table*}[t]
\setlength{\abovecaptionskip}{0.1cm}
\setlength{\belowcaptionskip}{0.1cm}
\centering
\begin{threeparttable}
\caption{Error (\%) of falsely rejecting genuine images (FRR) and accepting black-box adversarial examples (FAR) in FaceNet}
\label{tab_df_fn_sgadv_b}
\setlength{\tabcolsep}{4.2mm}{
\begin{tabular}{c|c|ccccc}
\hline
Defense&$FRR_{genuine}$&$FAR_{DI^2-FGSM}$&$FAR_{TI-FGSM}$&$FAR_{LGC}$&$FAR_{BIM}$&$FAR_{Square}$\\
\hline
\hline
FaceNet&1.20&55.73&54.93&52.63&50.97&100.0\\
\Xcline{1-7}{0.8pt}
DiffPure~\cite{nie2022DiffPure}&5.06&41.40&40.63&37.63&34.67&33.4\\
IWMF (ours)&6.38&\textbf{2.83}&\textbf{3.63}&\textbf{2.80}&\textbf{1.77}&30.4\\
IWMF-Diff (ours)&\textbf{3.80}&23.33&27.07&19.43&12.87&\textbf{27.6}\\
\hline
\end{tabular}}
\end{threeparttable}
\end{table*}

\begin{table*}[t]
\setlength{\abovecaptionskip}{0.1cm}
\setlength{\belowcaptionskip}{0.1cm}
\centering
\begin{threeparttable}
\caption{Error (\%) of falsely rejecting adversarial examples (FRR) as their TRUE identities in FaceNet}
\label{tab_rb_fn_sgadv_w}
\setlength{\tabcolsep}{5mm}{
\begin{tabular}{c|c|ccccc}
\hline
Defense&$FRR_{genuine}$&$FRR_{SGADV}$&$FRR_{FGSM}$&$FRR_{PGD}$&$FRR_{CW}$&$FRR_{APGD}$\\
\hline
\hline
FaceNet&1.20&99.48&33.52&74.54&63.86&98.22\\
\Xcline{1-7}{0.8pt}
PIN~\cite{ren2022perturbation}&17.60&17.86&17.86&18.02&15.90&19.44\\
DiffPure~\cite{nie2022DiffPure}&5.06&9.58&16.50&7.14&6.02&11.14\\
IWMF (ours)&6.38&12.18&11.32&9.30&7.88&14.06\\
IWMF-Diff(ours)&\textbf{3.80}&\textbf{7.26}&\textbf{7.64}&\textbf{5.40}&\textbf{4.98}&\textbf{7.22}\\
\hline
\end{tabular}}
\end{threeparttable}
\end{table*}

% \begin{table}[t]
% \setlength{\abovecaptionskip}{0.1cm}
% \setlength{\belowcaptionskip}{0.1cm}
% \centering
% \begin{threeparttable}
% \caption{Resistance (\%) to the adaptive attack for FaceNet.}
% \label{tab_counterattack_fn}
% \setlength{\tabcolsep}{1.8mm}{
% \begin{tabular}{c|c|cc}
% \hline
% Defense&$FAR_{imposter}$&$FAR_{SGADV}$&$FAR_{adaptive}$\\
% \hline
% \hline
% FaceNet&1.20&100.0&N/A\\
% \Xcline{1-4}{0.8pt}
% PIN~\cite{ren2022perturbation}&10.38&15.0&87.6\\
% DiffPure~\cite{nie2022DiffPure}&0.35&5.0&89.4\\
% IWMF (ours)&1.26&6.4&82.2\\
% IWMF-Diff (ours)&0.93&3.8&\textbf{70.8}\\
% \hline
% \end{tabular}}
% \end{threeparttable}
% \end{table}

\begin{table}[t]
\setlength{\abovecaptionskip}{0.1cm}
\setlength{\belowcaptionskip}{0.1cm}
\centering
\begin{threeparttable}
\caption{Resistance (\%) to the adaptive attack for FaceNet.}
\label{tab_counterattack_fn}
\setlength{\tabcolsep}{1.2mm}{
\begin{tabular}{c|c|cc}
\hline
Defense&$FAR_{SGADV}$&$FAR_{APGD-EOT}$&$FAR_{adaptive}$\\
\hline
\hline
FaceNet&100.0&100.0&N/A\\
\Xcline{1-4}{0.8pt}
PIN~\cite{ren2022perturbation}&15.0&19.6&87.6\\
DiffPure~\cite{nie2022DiffPure}&5.0&13.4&93.6\\
IWMF (ours)&6.4&9.2&84.2\\
IWMF-Diff (ours)&3.8&\textbf{8.4}&\textbf{77.8}\\
\hline
\end{tabular}}
% \begin{tablenotes}
% \footnotesize
% \item $^{*}$ denotes adaptively attacking the defense without randomization.
% \end{tablenotes}
\end{threeparttable}
\end{table}

\section{Ablation Study}
\label{ablation}
\subsection{Blurring and Diffusion Strategies}
\label{df_ablation_blur_diff}
IWMF is a proposed method for image blurring. Therefore, comparisons were made between IWMF and other blurring strategies. Specifically, the classic mean filter served as the backbone for IWMF, and its algorithm is referenced in Equation~\ref{eq_mean_filter}. Other strategies include median filter, pepper noise, Gaussian noise, iterative window median filter (IWMF with median computation), and non-iterative window mean filter (Figure~\ref{fig_iwmf}(e)). The samples of these blurring strategies are illustrated in Figure~\ref{fig_blur_diff}.

Regarding white-box adversarial attacks (such as SGADV), Table~\ref{tab_blur_if} shows that IWMF and IWMF-Diff are the best defense strategies for blurring and diffusion-based denoising, respectively. Specifically, \one~IWMF performs better than Gaussian noise; \two~the denoising diffusion model is ineffective in covering perturbations without blurring (referring to ``noiseless'' row); and \three~the combination of IWMF and Gaussian noise for blurring is superior to either individual blurring strategy, as the diffusion model is trained with Gaussian noise.

\begin{table}[t]
\setlength{\abovecaptionskip}{0.1cm}
\setlength{\belowcaptionskip}{0.1cm}
\centering
\begin{threeparttable}
\caption{EER (\%) of various blurring and diffusion strategies for InsightFace}
\label{tab_blur_if}
\setlength{\tabcolsep}{4mm}{
\begin{tabular}{ccc}
\hline
\multicolumn{2}{c}{Strategy}&SGADV\\
\hline
\hline
\multirow{7}{*}{Blurring}&median filter&57.40\\
&mean filter&33.40\\
&pepper noise&9.86\\
&Gaussian noise&7.60\\
&non-iterative window mean filter&8.34\\
&iterative window median filter&9.01\\
&IWMF (ours)&\textbf{6.36}\\
\hline
\multirow{4}{*}{Diffusion}&noisless (DDPM~\cite{ho2020denoising})&82.61\\
&IWMF&6.34\\
&Gaussian (DiffPure~\cite{nie2022DiffPure})&5.00\\
&IWMF+Gaussian (ours)&\textbf{3.22}\\
\hline
\end{tabular}}
\end{threeparttable}
\end{table}

\begin{figure*}[t]
\setlength{\abovecaptionskip}{0.1cm}
\setlength{\belowcaptionskip}{0.1cm}
\centering
\includegraphics[width=\linewidth]{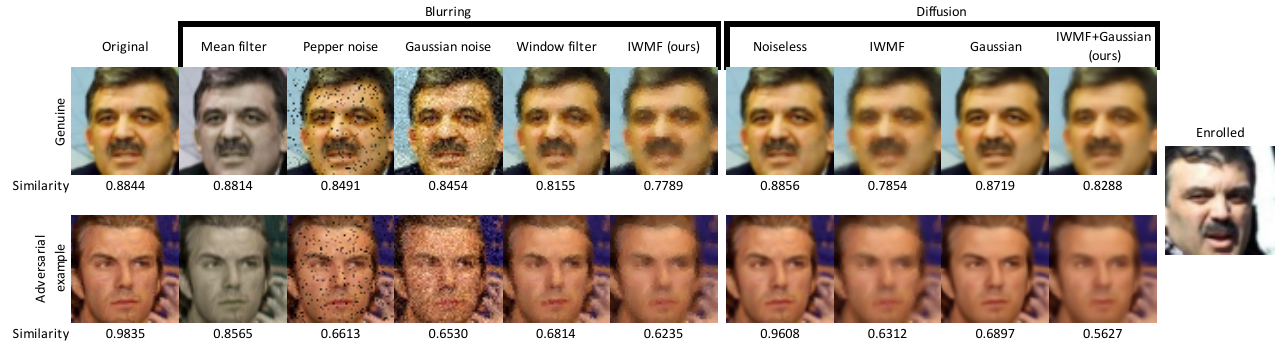}
\caption{Samples of various blurring and diffusion strategies.}
\label{fig_blur_diff}
\end{figure*}

\begin{figure}[t]
\setlength{\abovecaptionskip}{0.1cm}
\setlength{\belowcaptionskip}{0.1cm}
\centering
\includegraphics[width=\linewidth]{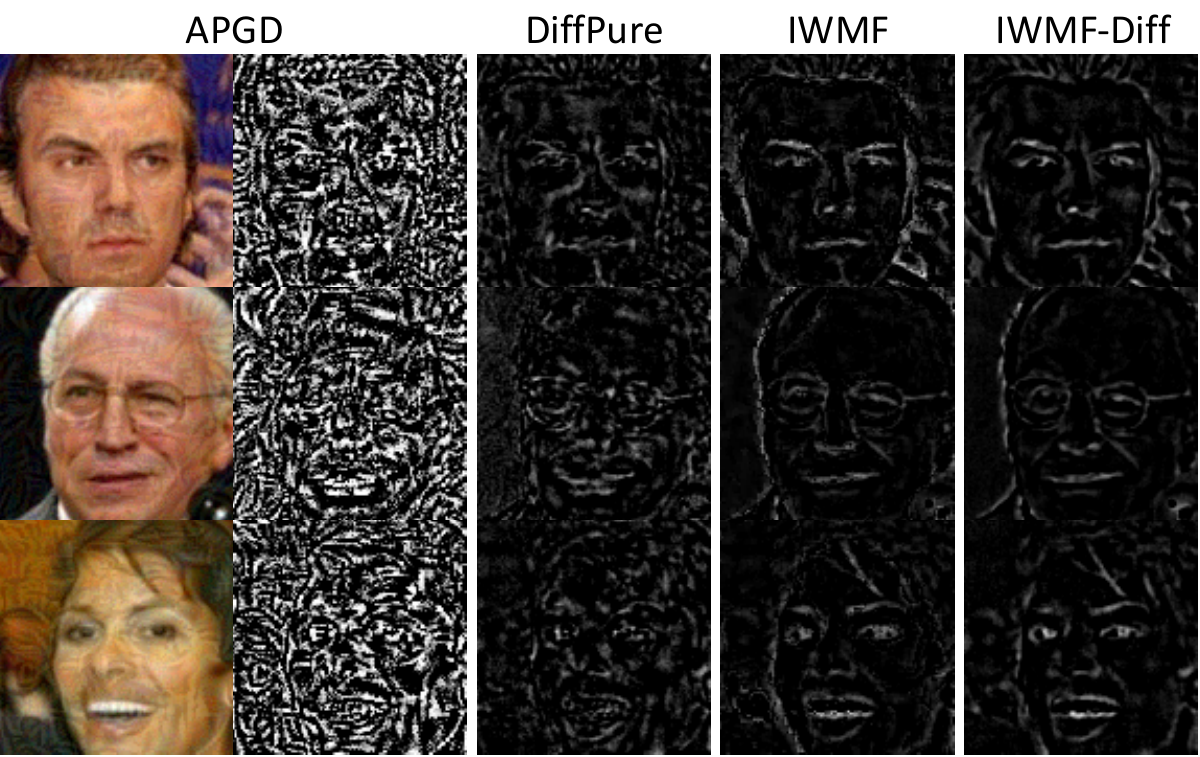}
\caption{The proposed IWMF and IWMF-Diff methods outperform DiffPure \cite{nie2022DiffPure} in purifying adversarial perturbations with respect to less visible perturbations. The perturbation size for APGD \cite{croce2020reliable} was set to 0.06, and the settings for the three purification methods can be found in Table \ref{tab_settings}. The target system used was InsightFace \cite{deng2019arcface}.}
\label{fig_perturbations}
\end{figure}

\begin{table*}[t]
\setlength{\abovecaptionskip}{0.1cm}
\setlength{\belowcaptionskip}{0.1cm}
\centering
\begin{threeparttable}
\caption{The effectiveness (\%) of the diffusion model against adversarial perturbations}
\label{tab_df_ablation_black}
\setlength{\tabcolsep}{4.4mm}{
\begin{tabular}{cccccccc}
\hline
Defense&$\lambda$&$\sigma_y$&$\tau$&$FRR_{genuine}$&$FAR_{SGADV}$&$FAR_{DI^2-FGSM}$&$FAR_{BIM}$\\
\hline
\hline
InsightFace&N/A&N/A&\multirow{4}{*}{0.6131}&0.28&100.0&95.00&91.50\\
\Xcline{1-3}{0.8pt}
\Xcline{5-8}{0.8pt}
DiffPure~\cite{nie2022DiffPure}&0&0.15&&\textbf{0.38}&69.4&94.77&89.80\\
IWMF (ours)&0.25&N/A&&0.78&64.6&\textbf{43.33}&\textbf{21.13}\\
IWMF-Diff (ours)&0.25&0.15&&1.44&\textbf{8.6}&44.30&22.30\\
\hline
\end{tabular}}
\end{threeparttable}
\end{table*}

Figure~\ref{fig_blur_diff} illustrates that visible image quality does not necessarily correlate with defense performance, as evidenced by the comparison between Gaussian-based diffusion and IWMF-Diff patches. This discrepancy suggests a difference in how deep learning models and humans interpret images, as discussed in Section~\ref{df_discuss_similarity}. Additionally, Figure~\ref{fig_perturbations} demonstrates that the proposed IWMF and IWMF-Diff more effectively purify adversarial perturbations.

However, the diffusion model is of limited help in covering perturbations from adversarial examples generated by black-box attacks, especially those based on facial landmarks. The results in Table~\ref{tab_df_ablation_black} were obtained under the same settings and evaluated against three representative adversarial attacks: SGADV (gradient-based white box), DI$^2$-FGSM (facial-landmark-based black box), and BIM (gradient-based black box). Specifically, when comparing the column labeled ``$FRR_{SGADV}$'', both the diffusion model and IWMF are effective against white-box attacks. When comparing the ``DiffPure'' and ``Insightface'' rows, it is evident that the diffusion model is unable to effectively cover perturbations from black-box attacks, as the FARs are negligibly reduced. This is further emphasized when comparing the ``IWMF'' and ``IWMF-Diff'' rows, as the FARs remain comparable even with $\sigma_y=0.15$. This suggests that the security enhancement achieved by the state-of-the-art defense DiffPure in Table~\ref{tab_df_if_sgadv_w} is mainly contributed by the threshold, rather than the defense itself. To defend against black-box attacks, it is recommended to use IWMF-based approaches. Additionally, compared between the columns labeled ``$FAR_{DI^2-FGSM}$'' and ``$FAR_{BIM}$'', gradient-based black-box adversarial attacks are easier to defend against than facial-landmark-based attacks.

\begin{figure*}[t]
\setlength{\abovecaptionskip}{0.1cm}
\setlength{\belowcaptionskip}{0.1cm}
\centering
\includegraphics[width=0.8\linewidth]{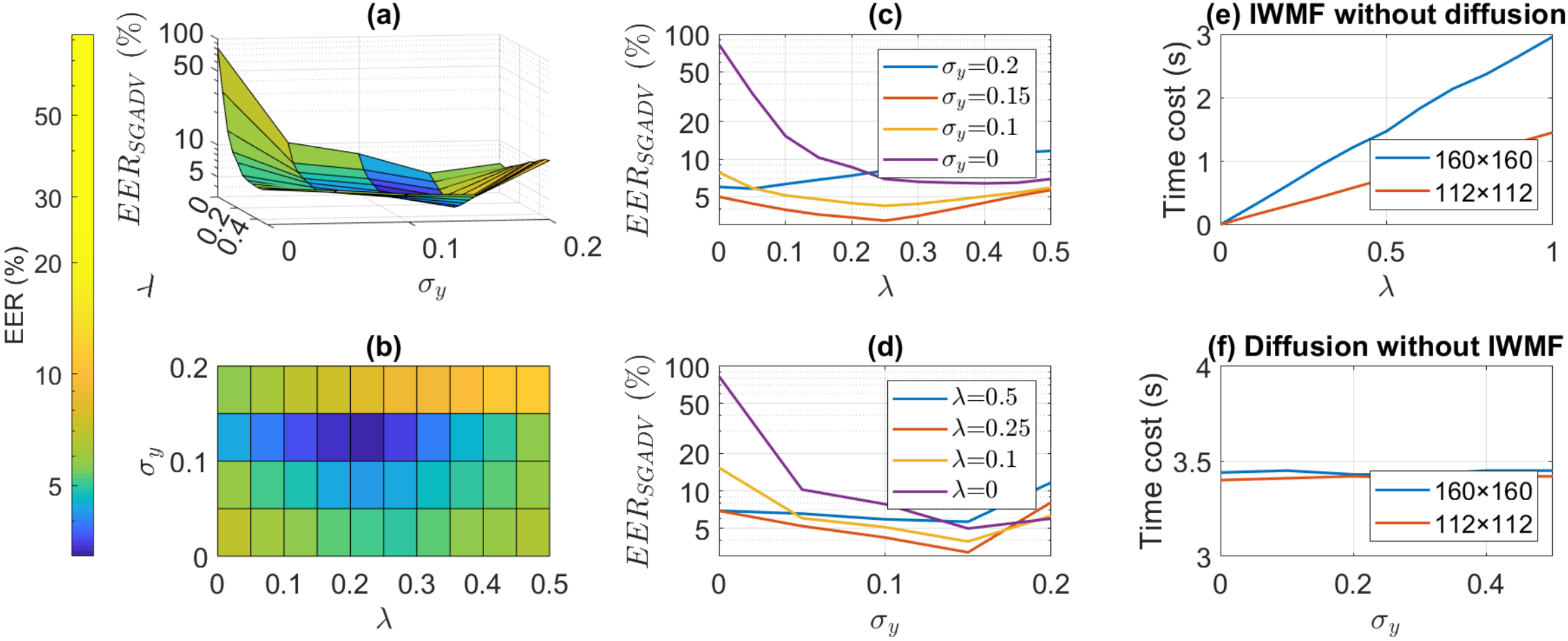}
\caption{Ablation studies on $\lambda$ and $\sigma_y$.}
\label{fig_ablation}
\end{figure*}

\subsection{Window amount and Gaussian Standard Deviation}
\label{df_ablation_3d}
As shown in Figures~\ref{fig_ablation}(a) and \ref{fig_ablation}(b), the defense performance depends on the values of window amount $\lambda$ and Gaussian standard deviation $\sigma_y$. The charts indicate that to achieve the smallest EER, neither $\lambda$ nor $\sigma_y$ should be too large, as this would indicate that blurring by IWMF and Gaussian noise could not be too heavy. This is a reasonable finding because if the blurring is too heavy, the genuine images become overly distorted, while if the blurring is too light, the adversarial examples are not effectively purified.

\begin{figure}[t]
\setlength{\abovecaptionskip}{0.1cm}
\setlength{\belowcaptionskip}{0.1cm}
\centering
\includegraphics[width=\linewidth]{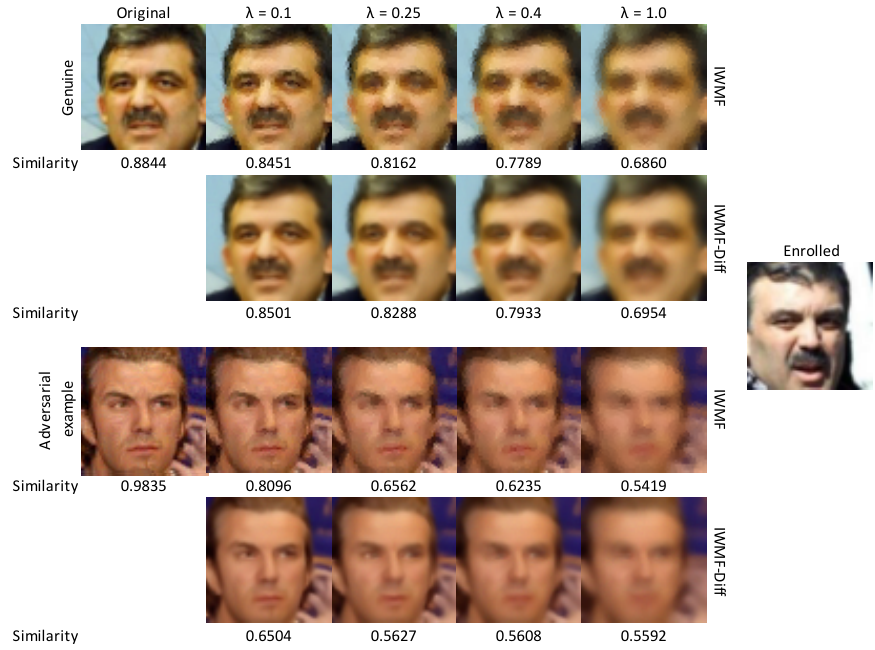}
\caption{Samples in various window amounts @ $\sigma_y=0.15$.}
\label{fig_lambda}
\end{figure}

\begin{figure}[!t]
\centering
\includegraphics[width=\linewidth]{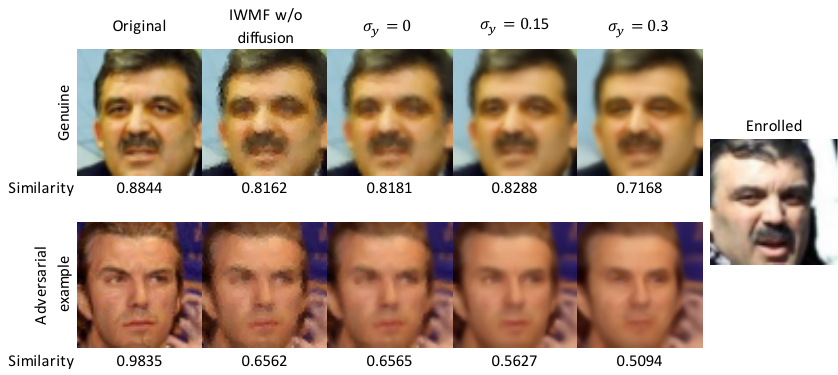}
\caption{Samples in various Gaussian standard deviation @ $\lambda=0.25$.}
\label{fig_sigma}
\end{figure}

Similar findings were observed in ablation studies on individual $\lambda$ and $\sigma_y$, as illustrated in Figures~\ref{fig_ablation}(c) and \ref{fig_ablation}(d). It is found that a larger $\lambda$ gives better performance when $\sigma_y$ is smaller. However, when either $\sigma_y$ or $\lambda$ was too large (\eg $\sigma_y=0.2$ or $\lambda=0.5$) or too small (\eg $\sigma_y=0$ or $\lambda=0$), the performance was not further improved. In terms of computational complexity, Figure~\ref{fig_ablation}(e) shows that a larger $\lambda$ results in more time cost, given the increased number of iterations (\ie windows) required by Step \ref{step_lambda} in Algorithm~\ref{alg_iwmf}. However, as demonstrated in Figure~\ref{fig_ablation}(f), the time cost of the diffusion-based denoising is not affected by changes in $\sigma_y$. Finally, the qualitative analysis in Figures~\ref{fig_lambda} and \ref{fig_sigma} indicates that an increase in $\lambda$ or $\sigma_y$ always results in heavier blurring and a smaller similarity score, both for genuine images and adversarial examples.

\subsection{DDRM's Restoration Strategies}
DDRM develops multiple image processing strategies, with three of them (denoising, super resolution, and deblurring) applicable to adversarial purification. These strategies were compared, and the results in Table~\ref{tab_restore_if} and Figure~\ref{fig_restore} demonstrate that the denoising strategy in IWMF-Diff provided the best performance, with the smallest EER achieved.

\begin{table}[t]
\setlength{\abovecaptionskip}{0.1cm}
\setlength{\belowcaptionskip}{0.1cm}
\centering
\begin{threeparttable}
\caption{EER (\%) of various DDRM's restoration strategies for InsightFace}
\label{tab_restore_if}
\setlength{\tabcolsep}{13mm}{
\begin{tabular}{cccc}
\hline
Strategy&SGADV\\
\hline
denoising (ours)&\textbf{3.22}\\
super resolution $\times$1&4.40\\
super resolution $\times$2&9.40\\
deblurring&5.60\\
\hline
\end{tabular}}
\end{threeparttable}
\end{table}

\begin{figure}[t]
\setlength{\abovecaptionskip}{0.1cm}
\setlength{\belowcaptionskip}{0.1cm}
\centering
\includegraphics[width=\linewidth]{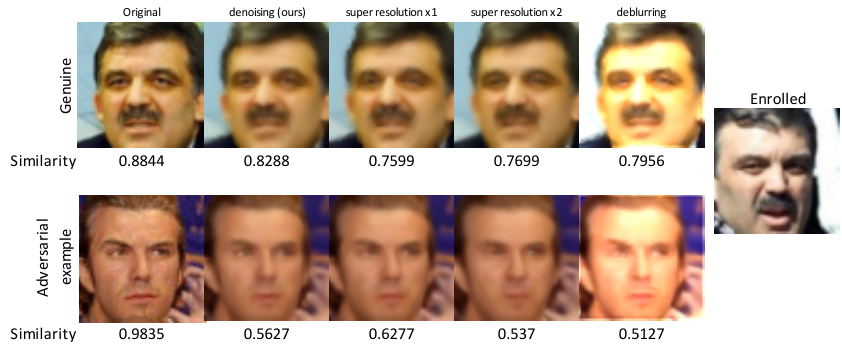}
\caption{Samples of various DDRM's restoration strategies.}
\label{fig_restore}
\end{figure}

\begin{table}[t]
\setlength{\abovecaptionskip}{0.1cm}
\setlength{\belowcaptionskip}{0.1cm}
\centering
\begin{threeparttable}
\caption{EER (\%) of IWMF-Diff in various window sizes}
\label{tab_ws_if}
\setlength{\tabcolsep}{15mm}{
\begin{tabular}{cccc}
\hline
Window size&SGADV\\
\hline
3px (ours)&\textbf{3.22}\\
5px&20.86\\
\hline
\end{tabular}}
\end{threeparttable}
\end{table}

\begin{figure}[t]
\setlength{\abovecaptionskip}{0.1cm}
\setlength{\belowcaptionskip}{0.1cm}
\centering
\includegraphics[width=0.7\linewidth]{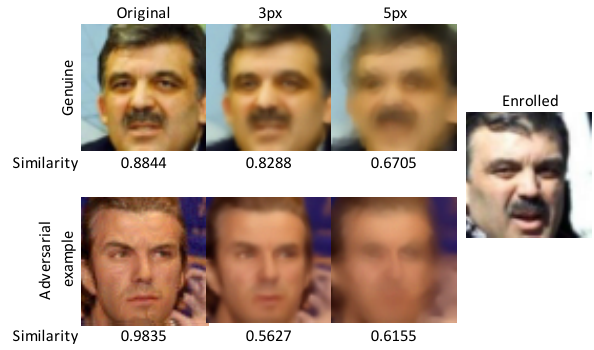}
\caption{Samples in various window sizes.}
\label{fig_window_size}
\end{figure}

\subsection{Window Size}
\label{window_size}
To distinguish IWMF from the classic mean filter, the minimum window size $s$ for IWMF was set to 3 pixels. However, as indicated in Table~\ref{tab_ws_if} and demonstrated in Figure~\ref{fig_window_size}, increasing the window size to $s=5$ pixels leads to a significant deterioration in performance. Therefore, the window size for IWMF and IWMF-Diff is fixed at 3 pixels.

\subsection{Threshold}
\label{df_ablation_thresh}
In authentication systems, one of the key factors that determines security and accuracy is the threshold. A larger threshold leads to better security (smaller FAR) but worse accuracy (larger FRR). To investigate the impact of the threshold on defense performance, two additional conditions were evaluated: $FRR_{genuine}=FAR_{imposter}$ and $FRR_{genuine}=FAR_{FGSM}$. The results in Tables \ref{tab_df_if_benign_w} and \ref{tab_df_if_fgsm_w} demonstrate that, as expected, the security of IWMF and IWMF-Diff can be improved by increasing the threshold, while accuracy can be increased by decreasing the threshold. Moreover, IWMF-Diff consistently outperforms other benchmark defenses across various thresholds.

\begin{table*}[t]
\setlength{\abovecaptionskip}{0.1cm}
\setlength{\belowcaptionskip}{0.1cm}
\centering
\begin{threeparttable}
\caption{Error (\%) of falsely rejecting genuine images (FRR) and accepting adversarial examples (FAR) in InsightFace @ $FRR_{genuine}=FAR_{imposter}$}
\label{tab_df_if_benign_w}
\setlength{\tabcolsep}{7.1mm}{
\begin{tabular}{cccccc}
\hline
Defense&$FRR_{genuine}$&$FAR_{SGADV}$&$FAR_{FGSM}$&$FAR_{PGD}$&$FAR_{CW}$\\
\hline
\hline
InsightFace&0.28&100.0&100.0&100.0&100.0\\
\Xcline{1-6}{0.8pt}
PIN~\cite{ren2022perturbation}&13.46&21.0&\textbf{20.4}&18.4&16.0\\
DiffPure~\cite{nie2022DiffPure}&\textbf{0.32}&74.2&98.4&69&6.8\\
IWMF (ours)&0.74&45.8&72.2&39.4&9.6\\
IWMF-Diff(ours)&0.72&\textbf{18.6}&50.8&\textbf{14.0}&\textbf{2.0}\\
\hline
\hline
\multirow{2}{*}{IWMF-Diff (fair)}&0.32&36.2&71.2&28.2&6.6\\
&2.60&4.2&20.4&2.0&0.2\\
\hline
\end{tabular}}
\end{threeparttable}
\end{table*}

\begin{table*}[t]
\setlength{\abovecaptionskip}{0.1cm}
\setlength{\belowcaptionskip}{0.1cm}
\centering
\begin{threeparttable}
\caption{Error (\%) of falsely rejecting genuine images (FRR) and accepting adversarial examples (FAR) in InsightFace @ $FRR_{genuine}=FAR_{FGSM}$}
\label{tab_df_if_fgsm_w}
\setlength{\tabcolsep}{7.1mm}{
\begin{tabular}{cccccc}
\hline
Defense&$FRR_{genuine}$&$FAR_{SGADV}$&$FAR_{FGSM}$&$FAR_{PGD}$&$FAR_{CW}$\\
\hline
\hline
InsightFace&0.28&100.0&100.0&100.0&100.0\\
\Xcline{1-6}{0.8pt}
PIN~\cite{ren2022perturbation}&16.20&13.8&16.2&17.0&11.2\\
DiffPure~\cite{nie2022DiffPure}&11.00&1.2&11.0&\textbf{0.0}&\textbf{0.0}\\
IWMF (ours)&9.94&3.8&9.8&0.2&\textbf{0.0}\\
IWMF-Diff(ours)&\textbf{7.10}&\textbf{0.6}&\textbf{7.0}&\textbf{0.0}&\textbf{0.0}\\
\hline
\end{tabular}}
\end{threeparttable}
\end{table*}

\subsection{Perturbation Size of Attacks}
\label{df_ablation_eps}
The size of the perturbation, denoted by $\epsilon$, in adversarial attacks is a crucial factor in determining the similarity between adversarial examples and their corresponding source images. Previous research has shown that larger values of $\epsilon$ lead to a higher attack success rate, increasing the difficulty of defending against such attacks. However, larger values of $\epsilon$ also make adversarial examples more easily detectable by the human eye. In the experiments conducted in previous sections, we adhered to the recommended values of $\epsilon$ provided in the respective attack papers to ensure practical image quality. In this section, we investigate the effect of varying $\epsilon$ on defense performance without considering image quality constraints. To this end, we set the value of $\epsilon$ to 0.5, 1, and 2 times the recommended value, as well as an extreme case where $\epsilon$ is set to its maximum value of 255/255.

The findings in Tables~\ref{tab_df_ablation_eps} and \ref{tab_df_ablation_eps_black} reveal that all defenses experience a decline in their ability to resist attacks as $\epsilon$ increases. However, both IWMF and IWMF-Diff remain more effective at larger values of $\epsilon$ compared to DiffPure, which exhibits FARs exceeding 90\% when $\epsilon$ is merely doubled. Additionally, although setting $\epsilon$ to 255/255 is impractical in real-world attacks, this setting represents the most theoretically challenging attacks. Nevertheless, the proposed defenses still demonstrate some resistance against these attacks, particularly in black-box scenarios.

\begin{table}[t]
\setlength{\abovecaptionskip}{0.1cm}
\setlength{\belowcaptionskip}{0.1cm}
\centering
\begin{threeparttable}
\caption{Resistance (\%) against attacks in different $\epsilon$}
\label{tab_df_ablation_eps}
\setlength{\tabcolsep}{3.2mm}{
\begin{tabular}{cccc}
\hline
Defense&$\epsilon$&$FAR_{APGD}$&$FAR_{Square}$\\
\hline
\hline
\multirow{4}{*}{InsightFace}&$\times$0.5&100&97.4\\
&$\times$1&100&100\\
&$\times$2&100&100\\
&255/255&100&100\\
\Xcline{1-4}{0.8pt}
\multirow{4}{*}{DiffPure~\cite{nie2022DiffPure}}&$\times$0.5&1.4&\textbf{0.4}\\
&$\times$1&17.4&20.4\\
&$\times$2&92.8&91.2\\
&255/255&100&79.0\\
\hline
\multirow{4}{*}{IWMF (ours)}&$\times$0.5&\textbf{0.6}&2.4\\
&$\times$1&9.2&28.8\\
&$\times$2&42.8&\textbf{64.8}\\
&255/255&\textbf{86.2}&\textbf{0.2}\\
\hline
\multirow{4}{*}{IWMF-Diff (ours)}&$\times$0.5&0.8&2.0\\
&$\times$1&\textbf{6.6}&\textbf{19.8}\\
&$\times$2&\textbf{37.2}&70.2\\
&255/255&95&7.0\\
\hline
\end{tabular}}
\begin{tablenotes}
\footnotesize
\item The setting of each defense refers to Table~\ref{tab_settings}.
\end{tablenotes}
\end{threeparttable}
\end{table}

\begin{table}[t]
\setlength{\abovecaptionskip}{0.1cm}
\setlength{\belowcaptionskip}{0.1cm}
\centering
\begin{threeparttable}
\caption{Effectiveness (\%) of the diffusion model against adversarial perturbations in different $\epsilon$}
\label{tab_df_ablation_eps_black}
\setlength{\tabcolsep}{3.2mm}{
\begin{tabular}{cccc}
\hline
Defense&$\epsilon$&$FAR_{APGD}$&$FAR_{Sqaure}$\\
\hline
\hline
\multirow{4}{*}{InsightFace}&$\times$0.5&100&97.4\\
&$\times$1&100&100\\
&$\times$2&100&100\\
&255/255&100&100\\
\Xcline{1-4}{0.8pt}
\multirow{4}{*}{DiffPure~\cite{nie2022DiffPure}}&$\times$0.5&41.8&33.2\\
&$\times$1&90.2&91.8\\
&$\times$2&100&99.8\\
&255/255&100&100\\
\hline
\multirow{4}{*}{IWMF (ours)}&$\times$0.5&9.4&20.0\\
&$\times$1&36.8&69.0\\
&$\times$2&76.4&90.2\\
&255/255&\textbf{97.4}&\textbf{4.6}\\
\hline
\multirow{4}{*}{IWMF-Diff (ours)}&$\times$0.5&\textbf{4.0}&\textbf{7.0}\\
&$\times$1&\textbf{14.6}&\textbf{39.4}\\
&$\times$2&\textbf{54.6}&\textbf{86.4}\\
&255/255&97.8&18.2\\
\hline
\end{tabular}}
\begin{tablenotes}
\footnotesize
\item The setting of each defense refers to Table~\ref{tab_df_ablation_black}.
\end{tablenotes}
\end{threeparttable}
\end{table}

\section{Discussion}
\label{discussion_df}

\subsection{Similarity after IWMF}
\label{df_discuss_similarity}
Our experiments in Figures~\ref{fig_lambda} and \ref{fig_window_size} demonstrate that the similarity between adversarial examples and the target images significantly decreases after applying IWMF blurring, particularly when using a larger $\lambda$ or window size $s$, as expected. However, we observed an intriguing result where, even when the genuine image is not visually recognizable (\eg using $s=5$px), the similarity remains higher than the threshold value of the original system (see Table~\ref{tab_settings}). This suggests that the deep learning model can still correctly extract the blurred facial features. As discussed in Section~\ref{iwmf}, the accuracy of feature extraction is preserved due to the IWMF blurring being confined within the ``neighbour'' distance, and not every pixel being changed. Additionally, our proposed iterative window filter's performance is superior, possibly due to its similarity with the operation of convolution kernels in deep learning models compared to other blurring strategies. This discovery uncovers an interesting research topic worth exploring further, \ie how deep learning models extract features from pixels.

\subsection{Denoising IWMF or Gaussian Noise}
\label{df_discuss_diff}
% As addressed in section \ref{df_ablation_blur_diff} and \ref{df_ablation_eps}, 1) IWMF outperforms other strategies for blurring including Gaussian noise; 2) Gaussian-based diffusion is infeasible to defend black-box attacks; 3) IMWF is more applicable to adversarial attacks with the larger $\epsilon$ than Gaussian-based diffusion. However, the images processed by IWMF cannot be restored by the diffusion model without additive Gaussian noise as all diffusion models are trained with Gaussian noise. This is the reason that IWMF-Diff adopts both IWMF and Gaussian noise for the best purification and restoration simultaneously. Nevertheless, the defense performance against white-box attacks is determined by IWMF and Gaussian noise together, yet for black-box attacks, only IWMF contributes majorly. It is reasonable to expect the performance further improved by training the denoising model with IWMF individually.

In Sections \ref{df_ablation_blur_diff} and \ref{df_ablation_eps}, we presented our findings on defense strategies against adversarial attacks. Our experiments revealed that \one~IWMF provides better blurring performance compared to other strategies, including Gaussian noise; \two~Gaussian-based diffusion is not effective against black-box attacks; and \three~IWMF shows superior defense performance against adversarial attacks with larger values of $\epsilon$ compared to Gaussian-based diffusion. 
However, one limitation of using IWMF blurring is that the images processed by IWMF cannot be restored accurately by the diffusion model without additive Gaussian noise. This is because all diffusion models are trained using Gaussian noise. This limitation led us to propose IWMF-Diff, which combines both IWMF and Gaussian noise to achieve the best purification and restoration simultaneously. It is worth noting that for white-box attacks, the defense performance is primarily determined by IWMF and Gaussian noise together, whereas for black-box attacks, only IWMF significantly contributes to the performance. Therefore, we suggest that training a denoising model with IWMF only could improve defense performance further.

\section{Conclusion}
This paper highlights critical defects in recent Gaussian-diffusion-based adversarial defenses. Specifically, we demonstrate that diffusion-based defenses suffer from efficiency issues and are not suitable for real-time applications. Moreover, Gaussian-diffusion-based adversarial purification is infeasible to defend black-box attacks, general attacks with large perturbations, and adaptive attacks. To address these challenges, we propose a novel, super-efficient, non-deep-learning-based image filter, called IWMF. Our experiments demonstrate that IWMF achieves comparable performance compared to state-of-the-art diffusion-based defenses and effectively alleviates the defects we identified. We also propose a pre-processing framework for adversarial purification, called IWMF-Diff, which is applicable to protect various deep learning models from different attack algorithms and outperforms the state-of-the-art defense. Furthermore, we evaluate the benchmark and our proposed defenses using the four requirements we define, which provide a comprehensive view of the defense performance. We believe that these four requirements can be useful for measuring newly proposed adversarial defenses. Our valuable discoveries regarding diffusion models and adversarial defenses can trigger a new research trend in this area.

For future works, as discussed in Section~\ref{discussion_df}, it is worth exploring the connection between IWMF and the operation of convolution kernels in deep learning models. Additionally, it would be interesting to train a new denoising model using IWMF to enhance defense performance further, and we recommend this as a potential topic for future research.

\footnotesize
%\nocite{*}
%\bibliographystyle{IEEEtran}
\bibliographystyle{plain}
\bibliography{sample-base}

\input{_bio}

\end{document}

%% file: _header.tex
\AtBeginDocument{%
  \providecommand\BibTeX{{%
    \normalfont B\kern-0.5em{\scshape i\kern-0.25em b}\kern-0.8em\TeX}}}

\usepackage[numbers,sort&compress]{natbib}
\usepackage{xcolor}
\usepackage{multirow}
\usepackage{makecell}
\usepackage[flushleft]{threeparttable}
\usepackage{algorithm}
\usepackage{algpseudocode}
\usepackage{graphicx}
\usepackage{url}
\usepackage{enumitem}
\usepackage{amsmath}
\usepackage{amssymb}
\usepackage{amsfonts}
\usepackage{xspace}
\def\eg{\emph{e.g.,}\xspace}

\def\ie{\emph{i.e.,}\xspace}
\def\etal{\emph{et al.}\xspace}

\newcommand{\one}{({\em i}\xspace)}
\newcommand{\two}{({\em ii}\xspace)}
\newcommand{\three}{({\em iii}\xspace)}
\newcommand{\four}{({\em iv}\xspace)}

\newcommand{\revise}[1]{\textcolor{blue}{#1}}

%% file: _authors.tex
%\thanks{Manuscript received April 19, 2021; revised August 16, 2021.}}
\author{
Hanrui Wang,~%\IEEEmembership{Member,~IEEE,}
Ruoxi Sun,~\IEEEmembership{Member,~IEEE,}
Cunjian Chen,~\IEEEmembership{Senior Member,~IEEE,}
Minhui Xue,~\IEEEmembership{Member,~IEEE,}
Lay-Ki Soon,~\IEEEmembership{Senior Member,~IEEE,}
Shuo Wang$^*$,~%\IEEEmembership{Fellow,~OSA,}
and 
Zhe Jin$^*$%~\IEEEmembership{Senior~Member,~IEEE,}% <-this % stops a space
\thanks{Hanrui Wang is with National Institute of Informatics (NII), Japan, e-mail: hanrui\_wang@nii.ac.jp. Ruoxi Sun, and Minhui Xue are with CSIRO's Data61, Australia, email: \{ruoxi.sun, jason.xue\}@data61.csiro.au. Cunjian Chen is with Monash University, Australia and Monash Suzhou Research Institute, China, e-mail: cunjian.chen@monash.edu. Lay-Ki Soon is with Monash University, Malaysia, email: soon.layki@monash.edu. Shuo Wang is with Shanghai Jiao Tong University, China, e-mail: wangshuosj@sjtu.edu.cn. Zhe Jin is with Anhui Provincial Key Laboratory of Secure Artificial Intelligence, Anhui University, China, email: jinzhe@ahu.edu.cn.}% <-this % stops a space
\thanks{This work was partially supported by JSPS KAKENHI Grants JP21H04907 and JP24H00732, by JST CREST Grants JPMJCR18A6 and JPMJCR20D3 including AIP challenge program, by JST AIP Acceleration Grant JPMJCR24U3, by JST K Program Grant JPMJKP24C2 Japan, and by the project for the development and demonstration of countermeasures against disinformation and misinformation on the Internet with the Ministry of Internal Affairs and Communications of Japan; the National Natural Science Foundation of China (Nos. 62376003) and Anhui Provincial Natural Science Foundation (No. 2308085MF200); the Faculty Initiatives Research, Monash University, via Contract No. 2901912, and support from the NVIDIA Academic Hardware Grant Program.}
\thanks{$^*$Corresponding authors: Zhe Jin and Shuo Wang.}
\thanks{Manuscript received April 7, 2023; revised August 20, 2024; accepted September 29, 2024.}
}
% \author{Hanrui Wang}
% \affiliation{%
%   \institution{Monash University}
%   %\streetaddress{1 Th{\o}rv{\"a}ld Circle}
%   \state{Selangor}
%   \country{Malaysia}}
% \email{hanrui.wang@monash.edu}

% \author{Shuo Wang}
% \affiliation{%
%   \institution{CSIRO}
%   \city{Sydney}
%   \country{Australia}
% }
% \email{shuo.wang@csiro.au}

% \author{Cunjian Chen}
% \affiliation{%
%  \institution{Monash University}
%  %\streetaddress{Rono-Hills}
%  \city{Suzhou}
%  %\state{Arunachal Pradesh}
% \email{cunjian.chen@monash.edu}

% \author{Massimo Tistarelli}
% \affiliation{%
%   \institution{University of Sassari}
%   %\streetaddress{30 Shuangqing Rd}
%   \city{Sassari}
%   %\state{Beijing Shi}
%   \country{Italy}}
% \email{tista@uniss.it}

% \author{Zhe Jin}
% %\authornote{Corresponding Author}
% \affiliation{%
%   \institution{Anhui University}
%   %\streetaddress{8600 Datapoint Drive}
%   \city{Hefei}
%   %\state{Texas}
%   \country{China}
%   %\postcode{78229}
% }
% \email{jinzhe@ahu.edu.cn}

% \author{Soon Lay Ki}
% \affiliation{%
%   \institution{Monash University}
%   %\streetaddress{1 Th{\o}rv{\"a}ld Circle}
%   \state{Selangor}
%   \country{Malaysia}}
% \email{soon.layki@monash.edu}

% %
% % By default, the full list of authors will be used in the page
% % headers. Often, this list is too long, and will overlap
% % other information printed in the page headers. This command allows
% % the author to define a more concise list
% % of authors' names for this purpose.
% \renewcommand{\shortauthors}{Wang, et al.}

%% file: _bio.tex
% Hanrui Wang
\begin{IEEEbiography}[{\includegraphics[width=1in,height=1.25in,clip,keepaspectratio]{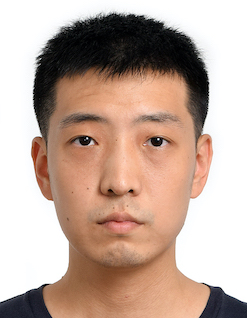}}]{Hanrui Wang}
received his B.S. degree in Electronic Information Engineering from Northeastern University (China) in 2011. He left the IT industry from a director position in 2019 to pursue a research career and received his Ph.D. in Computer Science from Monash University, Australia, in January 2024. He is currently working as a Postdoctoral Researcher with the Echizen Laboratory at the National Institute of Informatics (NII) in Tokyo, Japan. His research interests include AI security and privacy, particularly adversarial machine learning.
\end{IEEEbiography}

% Ruoxi Sun
\begin{IEEEbiography}[{\includegraphics[width=1in,height=1.25in,clip,keepaspectratio]{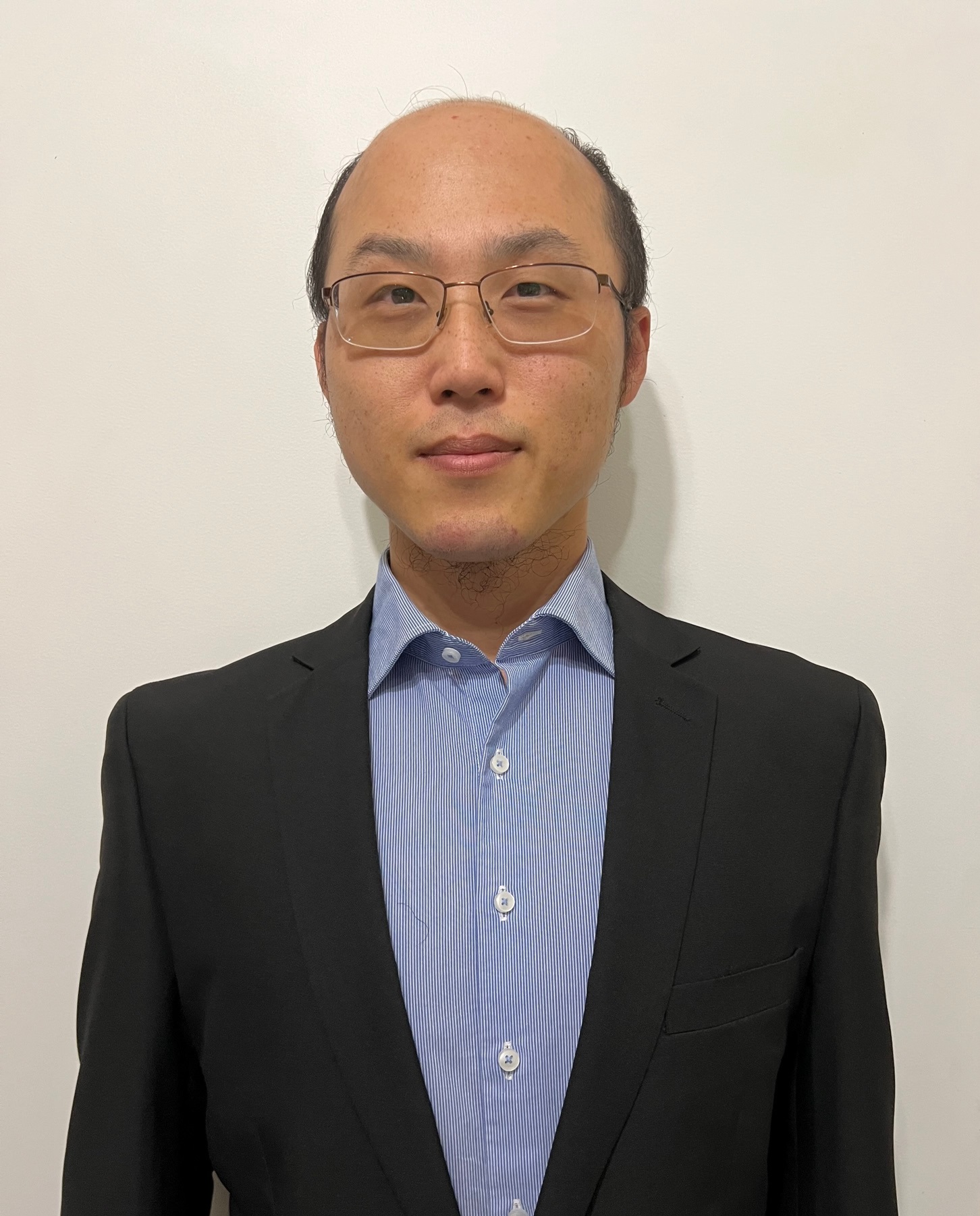}}]
{Ruoxi Sun} received Ph.D. degree from the University of Adelaide, Australia. He is currently a Research Fellow with CSIRO’s Data61, Australia. His research interests include mobile security and privacy, Internet of Things (IoT) security, and machine learning security. His work in the cybersecurity and privacy domains has led to the publication of over 25 papers in leading conferences and journals, such as the IEEE S\&P, ACM CCS, NDSS, WWW, ICSE, ACM FSE, and NeurIPS. Dr. Sun was a recipient of the ACM SIGSOFT Distinguished Paper Award. 
\end{IEEEbiography}

% Cunjian Chen
\begin{IEEEbiography}[{\includegraphics[width=1in,height=1.25in,clip,keepaspectratio]{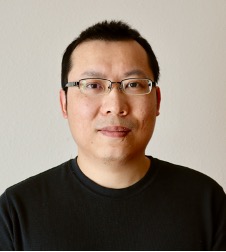}}]{Cunjian Chen}
is an Adjunct Lecturer at Monash University and a Research Fellow at the Monash Suzhou Research Institute. He previously worked as a Senior Research Associate at Michigan State University. He earned his PhD in Computer Science from West Virginia University and is currently an Associate Editor for Neural Processing Letters. He was also an Associate Editor for IET Image Processing. Dr. Chen has contributed to the academic community in various roles, including Tutorial Chair for IJCB, Area Chair for ICME, ICIP, ICASSP, and FG, and Session Chair for ICASSP, ICME, and FG. In recognition of his contributions, he received the Outstanding Area Chairs award at ICME 2021. He is also a Senior Member of IEEE.
\end{IEEEbiography}

% Jason Xue 
\begin{IEEEbiography}[{\includegraphics[width=1in,height=1.25in,clip,keepaspectratio]{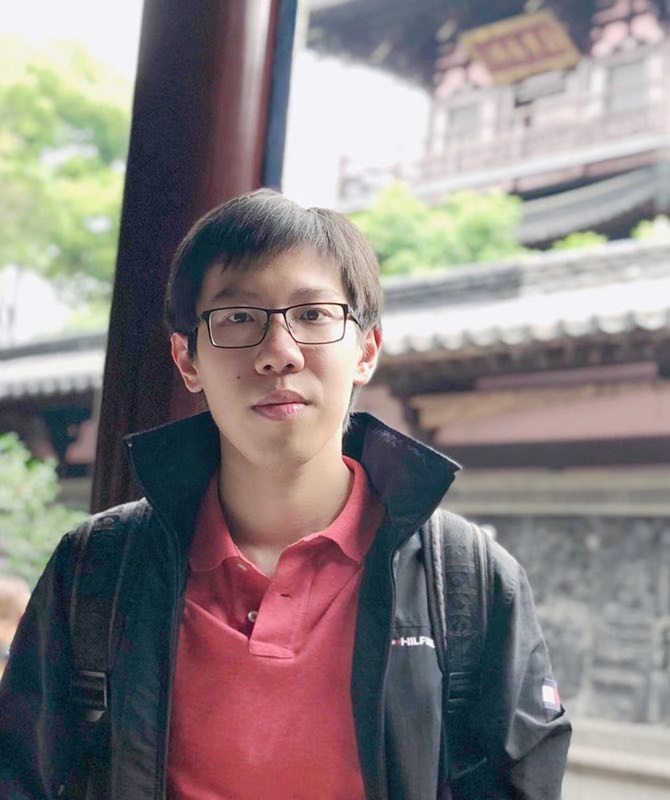}}]{Minhui Xue} is a Senior Research Scientist (lead of AI Security sub-team) at CSIRO's Data61, Australia. His current research interests are machine learning security and privacy, system and software security, and Internet measurement. He is the recipient of the ACM CCS Best Paper Award Runner-Up, ACM SIGSOFT distinguished paper award, Best Student Paper Award, and the IEEE best paper award, and his work has been featured in the mainstream press, including The New York Times, Science Daily, PR Newswire, Yahoo, The Australian Financial Review, and The Courier. He currently serves on the Program Committees of IEEE Symposium on Security and Privacy (Oakland) 2023, ACM CCS 2023, USENIX Security 2023, NDSS 2023, EuroS\&P 2023, ACM/IEEE ICSE 2023, and ACM/IEEE FSE 2023. He is a member of both ACM and IEEE.
\end{IEEEbiography}

% Soon Lay Ki
\begin{IEEEbiography}[{\includegraphics[width=1in,height=1.25in,clip,keepaspectratio]{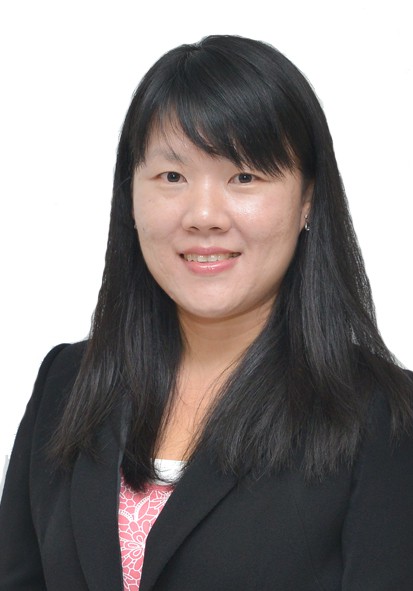}}]{Lay-Ki Soon} 
is an Associate Professor in School of Information Technology, Monash University Malaysia. She obtained her PhD from Soongsil University, South Korea. Her research interests include natural language processing, multimodal data analysis and data management. Lay-Ki has been awarded three competitive government grants as Lead Investigator. She is actively contributing to multidisciplinary research, such as emotion-aware chatbot for mitigating work anxiety, relation extractions from news articles and legal reasoning. In 2021, she was awarded the Faculty of IT Education Excellence Award in Citation for Outstanding Contribution to Student Learning. To date, she has graduated 8 PhD students and 6 Master students. Lay-Ki Soon is also a Senior Member of IEEE.
\end{IEEEbiography}

% Shuo Wang
\begin{IEEEbiography}[{\includegraphics[width=1in,height=1.25in,clip,keepaspectratio]{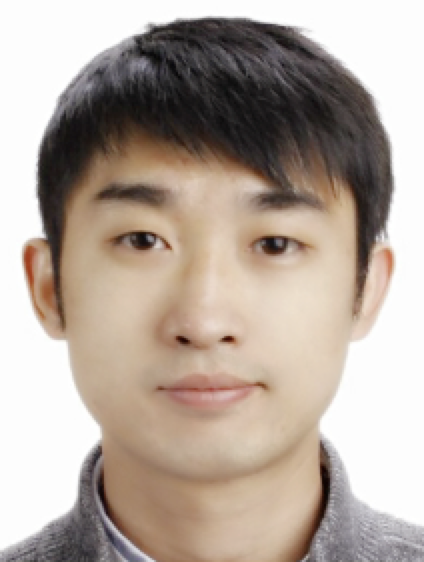}}]{Shuo Wang}
is an Associate Professor at Shanghai Jiao Tong University. Prior to this, He was a Senior Research Scientist at CSIRO, Australia's national science research agency. Shuo Wang's research endeavors are concentrated on the security implications within artificial intelligence and service systems. Shuo Wang has publications in IEEE S\&P, NDSS, USENIX Security, ICML, ICLR, TIFS, TDSC, TPDS, TSC, TNNLS, WWW, ESEC/FSE, and etc.
\end{IEEEbiography}

% Zhe Jin
\begin{IEEEbiography}[{\includegraphics[width=1in,height=1.25in,clip,keepaspectratio]{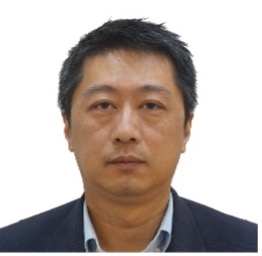}}]{Zhe Jin}
obtained the Ph.D. degree in engineering from University Tunku Abdul Rahman, Malaysia. Currently, he is a Professor at the School of Artificial Intelligence, Anhui University, China. His research interests include Biometrics, Pattern Recognition, Computer Vision, and Multimedia Security. He has received multiple highly competitive grants/projects, e.g., National Natural Science Foundation of China, Hundred Talents Project, and Anhui Provincial Natural Science Foundation with more than 1.2M Yuan. He has published more than 70 refereed journals and conference articles, including IEEE/ACM Trans., CVPR, NeurIPS and ICML. He was awarded Marie Skłodowska-Curie Research Exchange Fellowship at the University of Salzburg, Austria, and the University of Sassari, Italy, respectively, as a visiting scholar under the EU Project IDENTITY.
\end{IEEEbiography}

%% file: main.bbl
\begin{thebibliography}{10}

\bibitem{andriushchenko2020square}
Maksym Andriushchenko, Francesco Croce, Nicolas Flammarion, and Matthias Hein.
\newblock Square attack: a query-efficient black-box adversarial attack via
  random search.
\newblock In {\em European Conference on Computer Vision (ECCV)}, pages
  484--501, 2020.

\bibitem{athalye2018obfuscated}
Anish Athalye, Nicholas Carlini, and David Wagner.
\newblock Obfuscated gradients give a false sense of security: Circumventing
  defenses to adversarial examples.
\newblock In {\em International Conference on Machine Learning (ICML)}, pages
  274--283, 2018.

\bibitem{blau2022threat}
Tsachi Blau, Roy Ganz, Bahjat Kawar, Alex Bronstein, and Michael Elad.
\newblock Threat model-agnostic adversarial defense using diffusion models.
\newblock {\em arXiv preprint arXiv:2207.08089}, 2022.

\bibitem{Buc18}
Jacob Buckman, Aurko Roy, Colin Raffel, and Ian Goodfellow.
\newblock Thermometer encoding: One hot way to resist adversarial examples.
\newblock In {\em International Conference on Learning Representations (ICLR)},
  2018.

\bibitem{carlini2019evaluating}
Nicholas Carlini, Anish Athalye, Nicolas Papernot, Wieland Brendel, Jonas
  Rauber, Dimitris Tsipras, Ian Goodfellow, Aleksander Madry, and Alexey
  Kurakin.
\newblock On evaluating adversarial robustness.
\newblock {\em arXiv preprint arXiv:1902.06705}, 2019.

\bibitem{carlini2017adversarial}
Nicholas Carlini and David Wagner.
\newblock Adversarial examples are not easily detected: Bypassing ten detection
  methods.
\newblock In {\em Proceedings of the 10th ACM Workshop on Artificial
  Intelligence and Security (AISec)}, pages 3--14, 2017.

\bibitem{carlini2017towards}
Nicholas Carlini and David Wagner.
\newblock Towards evaluating the robustness of neural networks.
\newblock In {\em 2017 IEEE Symposium on Security and Privacy (S\&P)}, pages
  39--57, 2017.

\bibitem{chai2021ensembling}
Lucy Chai, Jun-Yan Zhu, Eli Shechtman, Phillip Isola, and Richard Zhang.
\newblock Ensembling with deep generative views.
\newblock In {\em Proceedings of the IEEE/CVF Conference on Computer Vision and
  Pattern Recognition (CVPR)}, pages 14997--15007, 2021.

\bibitem{cisse2017parseval}
Moustapha Cisse, Piotr Bojanowski, Edouard Grave, Yann Dauphin, and Nicolas
  Usunier.
\newblock Parseval networks: Improving robustness to adversarial examples.
\newblock In {\em International Conference on Machine Learning (ICML)}, pages
  854--863, 2017.

\bibitem{croce2021robustbench}
Francesco Croce, Maksym Andriushchenko, Vikash Sehwag, Edoardo Debenedetti,
  Nicolas Flammarion, Mung Chiang, Prateek Mittal, and Matthias Hein.
\newblock Robustbench: a standardized adversarial robustness benchmark.
\newblock In {\em Thirty-fifth Conference on Neural Information Processing
  Systems Datasets and Benchmarks Track (Round 2) (NeurIPS)}, 2021.

\bibitem{croce2022evaluating}
Francesco Croce, Sven Gowal, Thomas Brunner, Evan Shelhamer, Matthias Hein, and
  Taylan Cemgil.
\newblock Evaluating the adversarial robustness of adaptive test-time defenses.
\newblock In {\em International Conference on Machine Learning (ICML)}, pages
  4421--4435. PMLR, 2022.

\bibitem{croce2020reliable}
Francesco Croce and Matthias Hein.
\newblock Reliable evaluation of adversarial robustness with an ensemble of
  diverse parameter-free attacks.
\newblock In {\em International Conference on Machine Learning (ICML)}, pages
  2206--2216, 2020.

\bibitem{deng2019arcface}
Jiankang Deng, Jia Guo, Niannan Xue, and Stefanos Zafeiriou.
\newblock Arcface: Additive angular margin loss for deep face recognition.
\newblock In {\em Proceedings of the IEEE/CVF Conference on Computer Vision and
  Pattern Recognition (CVPR)}, pages 4690--4699, 2019.

\bibitem{dhillonstochastic}
Guneet~S Dhillon, Kamyar Azizzadenesheli, Zachary~C Lipton, Jeremy~D Bernstein,
  Jean Kossaifi, Aran Khanna, and Animashree Anandkumar.
\newblock Stochastic activation pruning for robust adversarial defense.
\newblock In {\em International Conference on Learning Representations (ICLR)},
  2018.

\bibitem{dong2019evading}
Yinpeng Dong, Tianyu Pang, Hang Su, and Jun Zhu.
\newblock Evading defenses to transferable adversarial examples by
  translation-invariant attacks.
\newblock In {\em Proceedings of the IEEE/CVF Conference on Computer Vision and
  Pattern Recognition (CVPR)}, pages 4312--4321, 2019.

\bibitem{feinman2017detecting}
Reuben Feinman, Ryan~R Curtin, Saurabh Shintre, and Andrew~B Gardner.
\newblock Detecting adversarial samples from artifacts.
\newblock {\em arXiv preprint arXiv:1703.00410}, 2017.

\bibitem{goodfellow2014explaining}
Ian~J Goodfellow, Jonathon Shlens, and Christian Szegedy.
\newblock Explaining and harnessing adversarial examples.
\newblock {\em arXiv preprint arXiv:1412.6572}, 2014.

\bibitem{graese2016assessing}
Abigail Graese, Andras Rozsa, and Terrance~E Boult.
\newblock Assessing threat of adversarial examples on deep neural networks.
\newblock In {\em 2016 15th IEEE International Conference on Machine Learning
  and Applications (ICMLA)}, pages 69--74, 2016.

\bibitem{gretton2012kernel}
Arthur Gretton, Karsten~M Borgwardt, Malte~J Rasch, Bernhard Sch{\"o}lkopf, and
  Alexander Smola.
\newblock A kernel two-sample test.
\newblock {\em The Journal of Machine Learning Research}, 13(1):723--773, 2012.

\bibitem{grosse2017statistical}
Kathrin Grosse, Praveen Manoharan, Nicolas Papernot, Michael Backes, and
  Patrick McDaniel.
\newblock On the (statistical) detection of adversarial examples.
\newblock {\em arXiv preprint arXiv:1702.06280}, 2017.

\bibitem{gu2014towards}
Shixiang Gu and Luca Rigazio.
\newblock Towards deep neural network architectures robust to adversarial
  examples.
\newblock {\em arXiv preprint arXiv:1412.5068}, 2014.

\bibitem{guo2018countering}
Chuan Guo, Mayank Rana, Moustapha Cisse, and Laurens van~der Maaten.
\newblock Countering adversarial images using input transformations.
\newblock In {\em International Conference on Learning Representations (ICLR)},
  2018.

\bibitem{hendrycks2016early}
Dan Hendrycks and Kevin Gimpel.
\newblock Early methods for detecting adversarial images.
\newblock {\em arXiv preprint arXiv:1608.00530}, 2016.

\bibitem{Hin15}
Geoffrey Hinton, Oriol Vinyals, and Jeff Dean.
\newblock Distilling the knowledge in a neural network.
\newblock {\em arXiv preprint arXiv:1503.02531}, 2015.

\bibitem{ho2020denoising}
Jonathan Ho, Ajay Jain, and Pieter Abbeel.
\newblock Denoising diffusion probabilistic models.
\newblock {\em Advances in Neural Information Processing Systems (NeurIPS)},
  33:6840--6851, 2020.

\bibitem{Hua08}
Gary~B Huang, Marwan Mattar, Tamara Berg, and Eric Learned-Miller.
\newblock Labeled faces in the wild: A database forstudying face recognition in
  unconstrained environments.
\newblock In {\em Workshop on Faces in 'Real-Life' Images: Detection,
  Alignment, and Recognition}, 2008.

\bibitem{Jain2011Introduction}
Anil~K. Jain, Arun~A. Ross, and Karthik Nandakumar.
\newblock {\em Introduction to Biometrics}.
\newblock Springer New York, NY, 2011.

\bibitem{jia2019comdefend}
Xiaojun Jia, Xingxing Wei, Xiaochun Cao, and Hassan Foroosh.
\newblock Comdefend: An efficient image compression model to defend adversarial
  examples.
\newblock In {\em Proceedings of the IEEE/CVF Conference on Computer Vision and
  Pattern Recognition (CVPR)}, pages 6084--6092, 2019.

\bibitem{jones1993lipschitzian}
Donald~R Jones, Cary~D Perttunen, and Bruce~E Stuckman.
\newblock Lipschitzian optimization without the lipschitz constant.
\newblock {\em Journal of optimization Theory and Applications}, 79:157--181,
  1993.

\bibitem{karras2020analyzing}
Tero Karras, Samuli Laine, Miika Aittala, Janne Hellsten, Jaakko Lehtinen, and
  Timo Aila.
\newblock Analyzing and improving the image quality of stylegan.
\newblock In {\em Proceedings of the IEEE/CVF Conference on Computer Vision and
  Pattern Recognition (CVPR)}, pages 8110--8119, 2020.

\bibitem{kawar2022denoising}
Bahjat Kawar, Michael Elad, Stefano Ermon, and Jiaming Song.
\newblock Denoising diffusion restoration models.
\newblock In {\em ICLR Workshop on Deep Generative Models for Highly Structured
  Data}, 2022.

\bibitem{kingma2013auto}
Diederik~P Kingma and Max Welling.
\newblock Auto-encoding variational bayes.
\newblock In {\em International Conference on Learning Representations (ICLR)},
  2014.

\bibitem{Kur16}
Alexey Kurakin, Ian~J Goodfellow, and Samy Bengio.
\newblock Adversarial machine learning at scale.
\newblock In {\em International Conference on Learning Representations (ICLR)},
  2016.

\bibitem{kurakin2018adversarial}
Alexey Kurakin, Ian~J Goodfellow, and Samy Bengio.
\newblock Adversarial examples in the physical world.
\newblock In {\em International Conference on Learning Representations (ICLR)
  Workshops}, pages 99--112, 2018.

\bibitem{laidlaw2021perceptual}
C~Laidlaw, S~Singla, and S~Feizi.
\newblock Perceptual adversarial robustness: Defense against unseen threat
  models.
\newblock In {\em International Conference on Learning Representations (ICLR)},
  2021.

\bibitem{lee2023robust}
Minjong Lee and Dongwoo Kim.
\newblock Robust evaluation of diffusion-based adversarial purification.
\newblock In {\em Proceedings of the IEEE/CVF International Conference on
  Computer Vision (ICCV)}, pages 134--144, 2023.

\bibitem{li2017adversarial}
Xin Li and Fuxin Li.
\newblock Adversarial examples detection in deep networks with convolutional
  filter statistics.
\newblock In {\em Proceedings of the IEEE International Conference on Computer
  Vision (ICCV)}, pages 5764--5772, 2017.

\bibitem{liao2018defense}
Fangzhou Liao, Ming Liang, Yinpeng Dong, Tianyu Pang, Xiaolin Hu, and Jun Zhu.
\newblock Defense against adversarial attacks using high-level representation
  guided denoiser.
\newblock In {\em Proceedings of the IEEE Conference on Computer Vision and
  Pattern Recognition (CVPR)}, pages 1778--1787, 2018.

\bibitem{Liu15}
Ziwei Liu, Ping Luo, Xiaogang Wang, and Xiaoou Tang.
\newblock Deep learning face attributes in the wild.
\newblock In {\em Proceedings of the IEEE International Conference on Computer
  Vision (ICCV)}, 2015.

\bibitem{madry2018towards}
Aleksander Madry, Aleksandar Makelov, Ludwig Schmidt, Dimitris Tsipras, and
  Adrian Vladu.
\newblock Towards deep learning models resistant to adversarial attacks.
\newblock In {\em International Conference on Learning Representations (ICLR)},
  2018.

\bibitem{meng2017magnet}
Dongyu Meng and Hao Chen.
\newblock Magnet: a two-pronged defense against adversarial examples.
\newblock In {\em Proceedings of the 2017 ACM SIGSAC Conference on Computer and
  Communications Security (ACM CCS)}, pages 135--147, 2017.

\bibitem{miller2020adversarial}
David~J Miller, Zhen Xiang, and George Kesidis.
\newblock Adversarial learning targeting deep neural network classification: A
  comprehensive review of defenses against attacks.
\newblock {\em Proceedings of the IEEE}, 108(3):402--433, 2020.

\bibitem{miyato2015distributional}
Takeru Miyato, Shin-ichi Maeda, Masanori Koyama, Ken Nakae, and Shin Ishii.
\newblock Distributional smoothing with virtual adversarial training.
\newblock {\em arXiv preprint arXiv:1507.00677}, 2015.

\bibitem{moosavi2016deepfool}
Seyed-Mohsen Moosavi-Dezfooli, Alhussein Fawzi, and Pascal Frossard.
\newblock Deepfool: a simple and accurate method to fool deep neural networks.
\newblock In {\em Proceedings of the IEEE Conference on Computer Vision and
  Pattern Recognition (CVPR)}, pages 2574--2582, 2016.

\bibitem{nie2022DiffPure}
Weili Nie, Brandon Guo, Yujia Huang, Chaowei Xiao, Arash Vahdat, and Anima
  Anandkumar.
\newblock Diffusion models for adversarial purification.
\newblock In {\em International Conference on Machine Learning (ICML)}, 2022.

\bibitem{papernot2017practical}
Nicolas Papernot, Patrick McDaniel, Ian Goodfellow, Somesh Jha, Z~Berkay Celik,
  and Ananthram Swami.
\newblock Practical black-box attacks against machine learning.
\newblock In {\em Proceedings of the 2017 ACM on Asia Conference on Computer
  and Communications Security (AsiaCCS)}, pages 506--519, 2017.

\bibitem{Pap16}
Nicolas Papernot, Patrick McDaniel, Xi~Wu, Somesh Jha, and Ananthram Swami.
\newblock Distillation as a defense to adversarial perturbations against deep
  neural networks.
\newblock In {\em 2016 IEEE symposium on security and privacy (S\&P)}, 2016.

\bibitem{pintor2022indicators}
Maura Pintor, Luca Demetrio, Angelo Sotgiu, Ambra Demontis, Nicholas Carlini,
  Battista Biggio, and Fabio Roli.
\newblock Indicators of attack failure: Debugging and improving optimization of
  adversarial examples.
\newblock {\em Advances in Neural Information Processing Systems (NeurIPS)},
  35:23063--23076, 2022.

\bibitem{ren2022perturbation}
Min Ren, Yuhao Zhu, Yunlong Wang, and Zhenan Sun.
\newblock Perturbation inactivation based adversarial defense for face
  recognition.
\newblock {\em IEEE Transactions on Information Forensics and Security},
  17:2947--2962, 2022.

\bibitem{richardson2021encoding}
Elad Richardson, Yuval Alaluf, Or~Patashnik, Yotam Nitzan, Yaniv Azar, Stav
  Shapiro, and Daniel Cohen-Or.
\newblock Encoding in style: a stylegan encoder for image-to-image translation.
\newblock In {\em Proceedings of the IEEE/CVF Conference on Computer Vision and
  Pattern Recognition (CVPR)}, pages 2287--2296, 2021.

\bibitem{samangouei2019defense}
Pouya Samangouei, Maya Kabkab, and Rama Chellappa.
\newblock Defense-gan: Protecting classifiers against adversarial attacks using
  generative models.
\newblock In {\em International Conference on Learning Representations (ICLR)},
  2019.

\bibitem{schroff2015facenet}
Florian Schroff, Dmitry Kalenichenko, and James Philbin.
\newblock Facenet: A unified embedding for face recognition and clustering.
\newblock In {\em Proceedings of the IEEE Conference on Computer Vision and
  Pattern Recognition (CVPR)}, pages 815--823, 2015.

\bibitem{shafahi2019adversarial}
Ali Shafahi, Mahyar Najibi, Mohammad~Amin Ghiasi, Zheng Xu, John Dickerson,
  Christoph Studer, Larry~S Davis, Gavin Taylor, and Tom Goldstein.
\newblock Adversarial training for free!
\newblock {\em Advances in Neural Information Processing Systems (NeurIPS)},
  32, 2019.

\bibitem{song2020denoising}
Jiaming Song, Chenlin Meng, and Stefano Ermon.
\newblock Denoising diffusion implicit models.
\newblock In {\em International Conference on Learning Representations (ICLR)},
  2020.

\bibitem{song2018pixeldefend}
Yang Song, Taesup Kim, Sebastian Nowozin, Stefano Ermon, and Nate Kushman.
\newblock Pixeldefend: Leveraging generative models to understand and defend
  against adversarial examples.
\newblock In {\em International Conference on Learning Representations (ICLR)},
  2018.

\bibitem{suciu2018does}
Octavian Suciu, Radu Marginean, Yigitcan Kaya, Hal Daume~III, and Tudor
  Dumitras.
\newblock When does machine learning $\{$FAIL$\}$? generalized transferability
  for evasion and poisoning attacks.
\newblock In {\em 27th USENIX Security Symposium (USENIX Security)}, pages
  1299--1316, 2018.

\bibitem{sun2022pointdp}
Jiachen Sun, Weili Nie, Zhiding Yu, Z~Morley Mao, and Chaowei Xiao.
\newblock Pointdp: Diffusion-driven purification against adversarial attacks on
  3d point cloud recognition.
\newblock {\em arXiv preprint arXiv:2208.09801}, 2022.

\bibitem{szegedy2014intriguing}
Christian Szegedy, Wojciech Zaremba, Ilya Sutskever, Joan Bruna, Dumitru Erhan,
  Ian Goodfellow, and Rob Fergus.
\newblock Intriguing properties of neural networks.
\newblock In {\em 2nd International Conference on Learning Representations
  (ICLR)}, 2014.

\bibitem{thys2019fooling}
Simen Thys, Wiebe Van~Ranst, and Toon Goedem{\'e}.
\newblock Fooling automated surveillance cameras: adversarial patches to attack
  person detection.
\newblock In {\em Proceedings of the IEEE/CVF Conference on Computer Vision and
  Pattern Recognition (CVPR) Workshops}, pages 0--0, 2019.

\bibitem{tramerensemble}
Florian Tram{\`e}r, Alexey Kurakin, Nicolas Papernot, Ian Goodfellow, Dan
  Boneh, and Patrick McDaniel.
\newblock Ensemble adversarial training: Attacks and defenses.
\newblock In {\em International Conference on Learning Representations (ICLR)},
  2018.

\bibitem{vahdat2020nvae}
Arash Vahdat and Jan Kautz.
\newblock Nvae: A deep hierarchical variational autoencoder.
\newblock {\em Advances in Neural Information Processing Systems (NeurIPS)},
  33:19667--19679, 2020.

\bibitem{wang2021similarity}
Hanrui Wang, Shuo Wang, Zhe Jin, Yandan Wang, Cunjian Chen, and Massimo
  Tistarelli.
\newblock Similarity-based gray-box adversarial attack against deep face
  recognition.
\newblock In {\em 2021 16th IEEE International Conference on Automatic Face and
  Gesture Recognition (FG 2021)}, pages 1--8, 2021.

\bibitem{wang2022guided}
Jinyi Wang, Zhaoyang Lyu, Dahua Lin, Bo~Dai, and Hongfei Fu.
\newblock Guided diffusion model for adversarial purification.
\newblock {\em arXiv preprint arXiv:2205.14969}, 2022.

\bibitem{wang2017adversary}
Qinglong Wang, Wenbo Guo, Kaixuan Zhang, Alexander~G Ororbia, Xinyu Xing, Xue
  Liu, and C~Lee Giles.
\newblock Adversary resistant deep neural networks with an application to
  malware detection.
\newblock In {\em Proceedings of the 23rd ACM SIGKDD International Conference
  on Knowledge Discovery and Data Mining (ACM SIGKDD)}, pages 1145--1153, 2017.

\bibitem{wu2022guided}
Quanlin Wu, Hang Ye, and Yuntian Gu.
\newblock Guided diffusion model for adversarial purification from random
  noise.
\newblock {\em arXiv preprint arXiv:2206.10875}, 2022.

\bibitem{xiemitigating}
Cihang Xie, Jianyu Wang, Zhishuai Zhang, Zhou Ren, and Alan Yuille.
\newblock Mitigating adversarial effects through randomization.
\newblock In {\em International Conference on Learning Representations (ICLR)},
  2017.

\bibitem{xie2019feature}
Cihang Xie, Yuxin Wu, Laurens van~der Maaten, Alan~L Yuille, and Kaiming He.
\newblock Feature denoising for improving adversarial robustness.
\newblock In {\em Proceedings of the IEEE/CVF Conference on Computer Vision and
  Pattern Recognition (CVPR)}, pages 501--509, 2019.

\bibitem{xie2019improving}
Cihang Xie, Zhishuai Zhang, Yuyin Zhou, Song Bai, Jianyu Wang, Zhou Ren, and
  Alan~L Yuille.
\newblock Improving transferability of adversarial examples with input
  diversity.
\newblock In {\em Proceedings of the IEEE/CVF Conference on Computer Vision and
  Pattern Recognition (CVPR)}, pages 2730--2739, 2019.

\bibitem{xu2020adversarial}
Han Xu, Yao Ma, Hao-Chen Liu, Debayan Deb, Hui Liu, Ji-Liang Tang, and Anil~K
  Jain.
\newblock Adversarial attacks and defenses in images, graphs and text: A
  review.
\newblock {\em International Journal of Automation and Computing}, 17:151--178,
  2020.

\bibitem{XuW171}
Weilin Xu, David Evans, and Yanjun Qi.
\newblock {Feature Squeezing: Detecting Adversarial Examples in Deep Neural
  Networks}.
\newblock In {\em Proceedings of the 2018 Network and Distributed Systems
  Security Symposium (NDSS)}, 2018.

\bibitem{yang2020robfr}
Xiao Yang, Dingcheng Yang, Yinpeng Dong, Hang Su, Wenjian Yu, and Jun Zhu.
\newblock Robfr: Benchmarking adversarial robustness on face recognition.
\newblock {\em arXiv preprint arXiv:2007.04118}, 2020.

\bibitem{zhang2019you}
Dinghuai Zhang, Tianyuan Zhang, Yiping Lu, Zhanxing Zhu, and Bin Dong.
\newblock You only propagate once: Accelerating adversarial training via
  maximal principle.
\newblock {\em Advances in Neural Information Processing Systems (NeurIPS)},
  32, 2019.

\bibitem{zhong2019adversarial}
Yaoyao Zhong and Weihong Deng.
\newblock Adversarial learning with margin-based triplet embedding
  regularization.
\newblock In {\em Proceedings of the IEEE/CVF International Conference on
  Computer Vision (ICCV)}, pages 6549--6558, 2019.

\bibitem{zhou2020manifold}
Jianli Zhou, Chao Liang, and Jun Chen.
\newblock Manifold projection for adversarial defense on face recognition.
\newblock In {\em European Conference on Computer Vision (ECCV)}, pages
  288--305, 2020.

\end{thebibliography}
